\documentclass[useAMS,usenatbib]{mn2e}
\usepackage{epsfig}
\usepackage{amsmath}
\usepackage{natbib}

\title[An optimized H$_{\beta}$ index for disentangling stellar population ages.]{An optimized H$_{\beta}$ index for disentangling stellar population ages}
\author[]{J.L. Cervantes$^{1}$\thanks{E-mail:
joseluis@iac.es} and A. Vazdekis$^{1}$\thanks{E-mail: vazdekis@iac.es }\\
$^{1}$Instituto de Astrof\'{\i}sica de Canarias, La Laguna, 38200 Tenerife, Spain\\
}
\begin{document}

\date{}

\pagerange{\pageref{firstpage}--\pageref{lastpage}} \pubyear{}

\maketitle

\label{firstpage}

\begin{abstract}

We have defined a new H$_{\beta}$ absorption index definition, H$_{\beta_{o}}$,
which has been optimized as an age indicator for old and intermediate age
stellar populations. Rather than using stellar spectra, we employed for this
purpose a library of stellar population SEDs of different ages and metallicities
at moderately high spectral resolution. H$_{\beta_{o}}$ provides us with
improved abilities for lifting the age-metallicity degeneracy affecting the
standard H$_{\beta}$ Lick index definition. The new index, which has also been
optimized against photon noise and velocity dispersion, is fully characterized
with wavelength shift, spectrum shape, dust extinction and [$\alpha$/Fe]
abundance ratio effects. H$_{\beta_{o}}$ requires spectra of similar qualites as
those commonly used for measuring the standard H$_{\beta}$ Lick index
definition. Aiming at illustrating the use and capabilities of H$_{\beta_{o}}$
as an age indicator we apply it to Milky Way globular clusters and to a well
selected sample of early-type galaxies covering a wide range in mass. The
results shown here are particularly useful for applying this index and
understand the involved uncertainties.

\end{abstract}



\section{Introduction}

To understand how galaxies form and evolve, we need to study their stellar 
populations as they are like fossils where the different formation and 
evolutionary processes are registered. Since stars are not resolved for  distant
stellar populations, one relies upon intergrated colors and spectra  to obtain
their physical parameters, such as ages or metallicities. However the integrated
light of galaxies suffers the well-known  age-metallicity degeneracy, making a
galaxy to look redder because it is older  or more metal rich (e.g,
\cite{1994ApJS...95..107W,1986A&A...164..260A}).

Unlike colors, spectroscopic absorption line-strength indices are more promising
at breaking the age-metallicity degeneracy, but even the most popular age
indicator, i.e., the Lick H$_{\beta}$ index (hereafter, H$_{\beta_{LICK}}$), 
does show a significant dependance on metallicity, particularly for old stellar 
populations \citep{1994ApJS...95..107W}. Despite the fact that other age
indicators based on  H$_{\gamma}$ feature have shown a larger sensitivity to age
than H$_{\beta_{LICK}}$ index, their signal-to-noise requirements are extremely
high or their dependence on spectral resolution and velocity dispersion make
them very difficult to apply for a large variety of data and targets
\citep{1995ApJ...446L..31J, 1999ApJ...525..144V}. 

Balmer lines are commonly used as age indicators, although they are not totally
immune to metallicity effects for integrated stellar populations. In order to
cope with the fact that H$_{\beta}$ can be filled in with nebular emission
\citep{1993PhDT.........7G}, indicators based on higher order Balmer lines
(H$_{\delta}$, H$_{\gamma}$)   have been proposed (e.g.,
\cite{1997ApJS..111..377W}, \cite{1999ApJ...525..144V}). The major drawback of
these index definitions based on the H$_{\delta}$ and H$_{\gamma}$ features is
that they have been shown to be significantly more sensitive to the total
metallicity and [$\alpha$/Fe] ratio than H$_{\beta_{LICK}}$
\citep{2004MNRAS.351L..19T,2005A&A...438..685K}. Nowadays the developement of
techniques where stellar and ionized-gas contributions to the galactic spectra
are simultaneously described allow us to decouple nebular emission from the
absorption features (e.g., \cite{2006MNRAS.366.1151S}).  Such procedures make it
possible to explote the potential of H$_{\beta}$ feature as an age indicator,
even when gas emission is present.  

Plotting H$_{\beta}$ index versus a metallicity indicator, measured on a set of
SSP spectra of different ages and metallicities, provides a diagnostic diagram
that allows us to partially lift the age-metallicity degeneracy. However it has
been shown the disadvantage of using H$_{\beta_{LICK}}$ versus various
metallicity indicators for measuring mean luminosity weighted ages of early-type
galaxies due to the fact that the resulting model grids are not fully
orthogonal. Indeed this method leads to younger age estimates when
H$_{\beta_{LICK}}$ is plotted versus Mg$_{b}$ than when is plotted versus an Fe
index if a galaxy is [Mg/Fe] overabundant (e.g., \cite{2006ApJ...637..200Y}). 
However a virtually orthogonal model grid is obtained when H$_{\beta_{LICK}}$ is
replaced by H$\gamma_{\sigma}$ \citep{1999ApJ...525..144V}. Unfortunately the
very high S/N required for measuring this index ( S/N (\AA) $>$ 150) limits its
applicability to nearby and bright objects for which such spectra can be
obtained. A more popular approach for obtaining consistent age estimates
requires the use of models that specifically take into account the non
scaled-solar elements ratios and the simultaneous measurement of several metal
lines to constrain the [$\alpha$/Fe] ratio
\citep{2004MNRAS.351L..19T,2005A&A...438..685K}. However, the obtained results
might depend on the details of modelling and on the particular element partition
employed. Furhermore the method is restricted to the use of the Lick/IDS system
of indices.  

In this paper, we explore, in a systematic manner, different index definitions
for the H$_{\beta}$ feature to find a new H$_{\beta}$ indicator that is
virtually insensitive to metallicity, which does not require particularly high
S/N. In section 2 we describe the optimization procedure developed to find the
new H$_{\beta}$ indicator, which makes use of a SSP library of SEDs at
moderately high resolution. In Section 3 we introduce the new H$_{\beta}$ index
definition and in section 4 we tackle its main characteristics. In section 5 we
test the reliability of this index on real data. Finally our conclusions are
summarized in section 6.


\begin{figure*}

\includegraphics[angle=0,height=8cm,width=8cm]{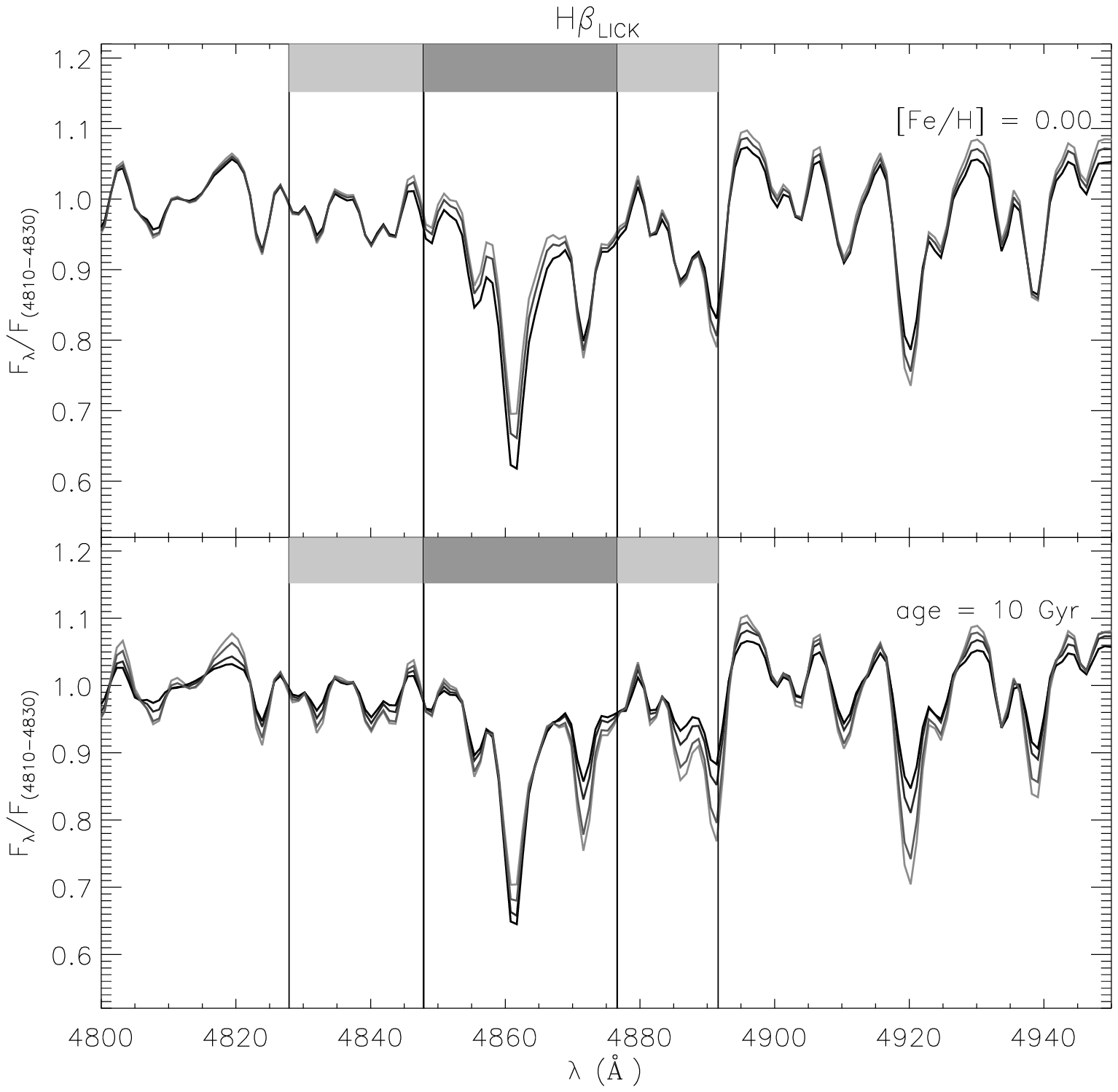}
\includegraphics[angle=0,height=8cm,width=8cm]{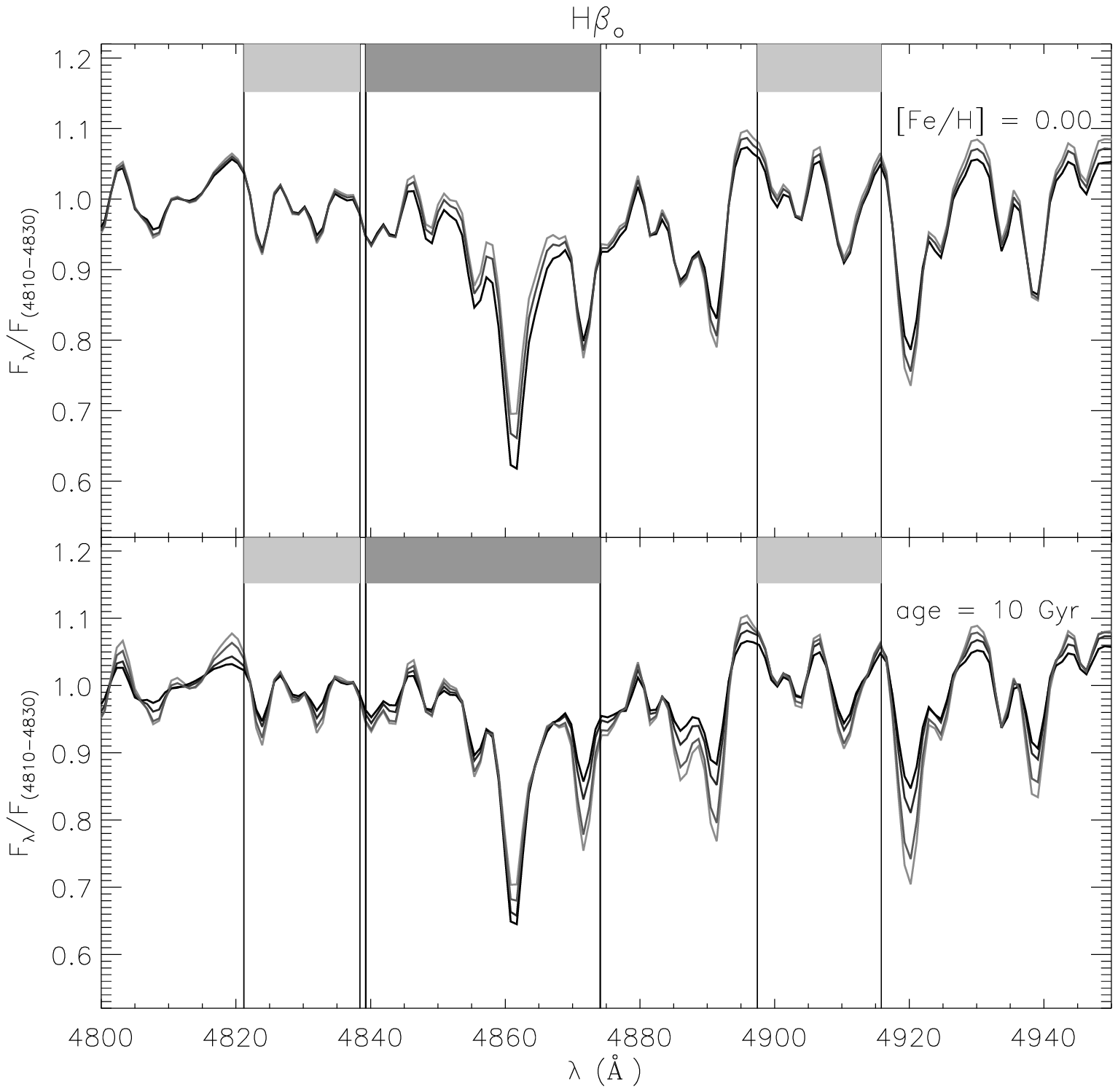}
\caption{The bandpasses of H$_{\beta_{LICK}}$ (\emph{left panel}) and
H$_{\beta_{o}}$ (\emph{right panel}) index definitions are shown on the top of
the panels, which include several SSP models.  \textsf{Top panels}: Solar
metallicity SSP model SEDs of age = 3, 7 and 12\,Gyr (from black to light grey,
respectively). \textsf{Bottom panels}: SSP models of age 10\,Gyr and
metallicities [Fe/H] = +0.2, +0.0, -0.4 and -0.7 (from light grey to black,
respectively). The improvement of the H$_{\beta_{o}}$ definition is achieved by
avoiding the metal lines on the red-pseudocontinnum, which are included in the
H$_{\beta_{LICK}}$ definition.} 
\label{figure1}
\end{figure*}

\section[]{Index definition approach}

In the Lick system
\citep{1984ApJ...287..586B,1993ApJS...86..153G,1994ApJS...94..687W,1997ApJS..111..377W},
an index is defined in terms of a central bandpass enclosing the feature
bracketed by two pseudocontinuum bandpasses at either side of the feature (blue
and red). Once the average fluxes in the pseudocontinua are obtained, a line is
drawn between their midpoints to represent the continuum of the central feature
bandpass allowing to define an index as a pseudoequivalent width.

The standard Lick/IDS system of indices was defined on the basis of a stellar
spectral library that were not flux-calibrated and had a resolution $\sim$ 8.4
{\AA} FWHM (three times lower than achieved in modern galaxy surveys, such as
SDSS). Furthermore, the Lick/IDS spectral resolution is varying with $\lambda$ 
\citep{1997ApJS..111..377W}.  The natural resolution of a galaxy spectrum is
given by the  convolution of the employed instrumental resolution with the
velocity  broadening due to galaxy dynamics. Therefore to apply any model
predictions based on the Lick/IDS system we need to smooth higher resolution 
spectra of galaxies with velocity dispersion values lower than that 
corresponding to the Lick/IDS system to match its resolution  (i.e. $\sigma \geq
{\rm 200\,km\,s^{-1}}$).  Furthermore to apply these model predictions to higher
velocity dispersion  galaxies we also need to correct index measurements back to
the resolution  of the Lick/IDS system (an uncertain method that usually
requires observing  a set of reference Lick stars.  Alternatively, we prefer to
use here SSP SEDs at higher resolution that  allow us to measure absorption line
indices directly on the model spectra  once smoothed to match galaxy velocity
dispersion. This allows a direct  comparison to the index values measured on the
galaxy spectrum.  This approach is advantageous for all stellar population
systems,  no matter its velocity dispersion nor its instrumental resolution.

We explote here the advantadge of the SSP SEDs to derive new indices or to
redefine previous ones, rather than employing stellar spectra or polynomial
fitting functions that relate the strengths of the absorption lines to stellar
atmospheric parameters. This new generation of stellar population synthesis
models allows us to perfom such analysis directly on the SSP spectra, allowing
us to understand how sensitive to the physical parameters of the stellar
populations a trial index definition is, thus avoiding the intermediate step of
modeling this index for individual stars. Here we accomplish this approach by
using the library of single-age, single-metallicity stellar population model
spectra of \citet{1999ApJ...513..224V} (hereafter V99) and its recent extension
\citet{vazdekis} (V07) on the basis of the stellar library MILES
(\cite{2006MNRAS.371..703S}). The fact that MILES is characterized by an
unprecedented stellar atmospheric parameter coverage
\citep{2007MNRAS.374..664C} allows predicting scaled-solar SEDs for old and
intermediate ages with metallicities $-1.7 \leq [M/H] \leq +0.2$ for the full
optical spectral range (3540-7410 {\AA}).

The use of SSP SEDs to define new indicators does not only allow us to optimize
them to be sensitive to the main population parameters such as age or
metallicity, but also to minimize their sensitivity, for example, to the
instrumental resolution or S/N requirements. This is possible as we can 
simulate spectra of stellar populations for different S/N and/or resolution (or
velocity dispersion) values. Furthermore with these SSP spectra it is
straightforward to simulate and analyze the effects of radial velocity (or
rotation curve), spectrum shape and dust extinction on each trial index
definition. 

For finding a new H$_{\beta}$ indicator we have adopted a Lick style index
definition, which consists of three bandpasses (feature and two
pseudocontinua). We then change the position of these bandpasses or modify
their widths until optimizing its sensitivity to a given parameter. Note that
there are alternative approaches for defining indices such as those of the Rose
system \citep{1985AJ.....90.1927R}, which are defined as the ratio between the
intensity of the central peak of two spectral features. Other index definitions
are for example the \emph{generic indices} \citep{2001MNRAS.326..959C} or those
considered pseudo-colors, e.g. D4000 \citep{1983ApJ...273..105B}. 

\subsection{Optimizing criteria}

The goal of this work is to define a new index based on H$_{\beta}$ feature
suitable for obtaining accurate age estimates.  We have carried out a
comprehensive analysis of this feature in order to obtain an index that is
virtually insensitive to the total metallicity, with great stability against
the smearing of this feature due to galaxy velocity dispersion (or instrumental
resolution) and with as lower S/N requirement as possible. Our method to derive
an optimized H$_{\beta}$ index definition developes a multicriteria analysis to
evaluate the above requirements for each trial index definition.

\subsubsection{Sensitivity to age and metallicity}

Our primary criterion is obtaining an index definition for the H$_{\beta}$
feature that maximizes our ability to measure ages. To investigate whether
metallicity effects can be decreased in the H$_{\beta}$ feature, we have
generalized the sensitivity parameter defined in \citet{1994ApJS...94..687W}
(see in there footnote 4):

\begin{equation}
\frac{\beta}{\alpha} = \frac{\sqrt{\sum_{i} \Big[\frac{(\frac{1}{I_{i}})((\frac{\partial I_{i}}{\partial [M/H]})_{age}) }
{(\frac{1}{I_{i}})((\frac{\partial I_{i}}{\partial log(age)})_{[M/H]}) } \Big]^{2}}}{N}
\end{equation}

\noindent where $i$ runs for SSP models with ages older than  5 \,Gyr and -0.7
$\leq [M/H] \leq$ +0.2, $i = 1,N$, and $I_{i}$ is the index value measured on
the $i$ SSP model for a  given index definition and $N$ is the number of
employed SSPs. Metallicity indicators should provide values above 1, whereas
the value for an age indicator should tend to  0. \looseness-1

Among the sistematically-generated configuration of
bandpasses, our method consists in choosing the ones that provide mininum
values for the $\frac{\beta}{\alpha}$ parameter. Note that the choice of
bandpasses that minimazes $\frac{\beta}{\alpha}$ should not depend on the
stellar population synthesis models in use. 

The parameter space where $\frac{\beta}{\alpha}$ is minimized covers the ages
and metallicities for most early-type galaxies \citep{1998yCat..21160001T} and,
for the H$_{\beta}$ case, represents the parameter subspace where the age
sensitivity is lower and where the metallicity  has the larger effects. In
other words, we are optimizing our indicator to be sensitive to the age for the
worst possible cases. 

\subsubsection{Spectral resolution and velocity dispersion}

One of the metioned advantages of using the SSP SEDs at higher resolution is
that it is straightforward to study the effects of the velocity dispersion (or
resolution) on a given line index definition. The models can be smoothed to
different levels, allowing us to evaluate its stability to $\sigma$ variations.
Note that $\sigma$ effects might be significant for some index definitions that
already are optimal according to the $\frac{\beta}{\alpha}$ parameter, as a
sufficiently small $\sigma$ variation might decrease the obtained age
sensitivities. To parametrize this effect, we calculate the partial derivative
of the index versus $\sigma$ for a solar metallicity SSP model of 10\,Gyr at
$\sigma = {\rm 150\,km\,s^{-1}}$:

\begin{equation}
\Sigma = \frac{1}{I} \cdot\left(\frac{\partial I}{\partial \sigma}\right)_{\rm{[M/H]}=0.0, T=10\,Gyr \arrowvert_{\sigma = 150 kms^{-1}}} 
\end{equation}

In a multicriteria analysis, the parameters that have been taken into account
are not assigned the same weight, as our main purpose is to optimize our age
resolving power. If we were to obtain an index definition that provides the
largest $\sigma$-stability, we would not have retained those definitions
providing the better age disentangling sensitivity. Therefore we have discarded
index definitions with $ \Sigma > {\rm 4 \cdot 10^{-4}}$ within the subset of
solutions for which the $\frac{\beta}{\alpha}$ parameter was optimized. This is
equivalent to a maximum variation of 10\% when comparing the index value at the
nominal resolution of the models and at ${\rm 250\,km\,s^{-1}}$.

\begin{table*}
 \centering
 \begin{minipage}{140mm}
  \caption{H$_{\beta}$ INDEX DEFINITIONS }\label{table1}
  \begin{tabular}{@{}lrrrrr@{}}
  \hline
{}    &  {Blue pseudocontinuum}   &
{Feature}  &
{Red pseudocontinuum}&{}&{}  \\
\cline{2-6} \\
{Index} & {\AA} & {\AA} &
 {\AA}&{c$_{1}$}& {c$_{2}$} \\
  \hline
 H$_{\beta_{LICK}}$& 4827.875  4847.875  & 4847.875  4876.625&  4876.625 
4891.625&7.301& 0.2539\\
H$_{\beta_{o}}$& 4821.175  4838.404  & 4839.275  4877.097&  4897.445 
4915.845&9.025 & 0.2386\\

\hline
\end{tabular}
\end{minipage}
\end{table*}

\subsubsection{Signal-to-Noise requirements}

We have studied the S/N requirements following the analytical  approach of 
\cite{1999PhDT........12C} (eqs. [9] and [43]-[44]) \footnote{Coefficients c1
and c2 are calculated from their equations (43) and (44) using definitions in
Table \ref{table1} and added there.}. Obviuously no index definition is
appropriate if the required S/N is extremely high. However, achieving minimum
index errors does not imply a mininum age uncertainty if the age indicator is
not totally insensitive to metallicity effects.  In fact, the age-metallicity
degeneracy and the uncertainity of age derived from index errors are tightly
correlated \citep{1998yCat..21160001T}. We consider that
the indicator depends solely on age, so that the uncertanity on measuring
this parameter comes mostly from photon noise. Aiming at obtaining index
definitions that minimize the S/N requirement, we define $\sigma[I_{a}]$ as the
index error required to achieve a precission of 2.5\,Gyr on deriving the age for a
a 10\,Gyr and solar metallicity model. Once again we relax our S/N optimizing
criterium by discarding only those definitions whose S/N(\AA) requirements are
above 70. 


\section{The new H$_{\beta}$ index }

According to our method for defining indices we show in this section an
optimized age indicator based on the H$_{\beta}$ feature, hereafter
H$_{\beta_{o}}$. Table \ref{table1} lists the limiting wavelength of the
bandpasses for the H$_{\beta_{LICK}}$ and H$_{\beta_{o}}$ indices. The
bandpasses of two indices are shown in Figure \ref{figure1}. Although farther
along we show plots and results obtained on the basis of V07, virtually
identical results are achieved with V99 models.

H$_{\beta_{LICK}}$ shows little sensitivity to the metallicity,  as no strong
metallic lines are included in the wavelength  range of the  index definition
\citep{2005A&A...438..685K}. However this dependence is not negligible as
inferred from the non-orthogonal model grids resulting when this index is
plotted versus the metallicity indicator [MgFe],  \citep{2006ApJ...637..200Y}.
Furthermore this prevents us to derive a unique age, when different
metallicity  indicators are used if the analyzed galaxies do not have
scaled-solar abundance patterns, particularly for old galaxies (see, for
example, their Figure \ref{figure7}).  Metallic  lines within H$_{\beta}$ are
mainly dominated by Mg via MgH absorption \citep{1995AJ....110.3035T}, Ti
\citep{2004MNRAS.353..917T} and Cr at $\sim$ 4885 {\AA}
\citep{2004MNRAS.351L..19T,2005A&A...438..685K}.

H$_{\beta_{o}}$ index definition avoids the spectral range covering those
metallic lines as the red-pseudocontinuum is shifted toward the red, whereas
the blue-pseudocontinuum is narrower to decrease the metallicity dependency.
However, other metallicity variations affect the H$_{\beta_{LICK}}$ central
bandpass via a Ti line at 4871\,{\AA} and the H$_{\beta}$ line at $\sim$ 4862
{\AA}. Note that the depth of both lines show an opposed behaviour against
metallicity: the higher the metallicity, the larger is the strength of the Ti 
line and the smaller the H$_{\beta}$ line. These opposed responses at
increasing metallicity are not balanced by each other within H$_{\beta_{LICK}}$
index definition. This index compesates this effect in part by incorporating a
metallicity dependence on the red-pseudocontinuum. In fact, it overcompensates
the global effect leading to a higher metallicity sensitivity. H$_{\beta_{o}}$
extends to the blue the blue-wing of the central bandpass to introduce the
required metallicity dependence, thus avoiding to introduce Cr in the
red-pseudocontinnuum, which causes the variation of H$_{\beta_{LICK}}$ with the
metallicity (Figure \ref{figure1}).  

\begin{figure*}
\includegraphics[angle=0,scale=.50]{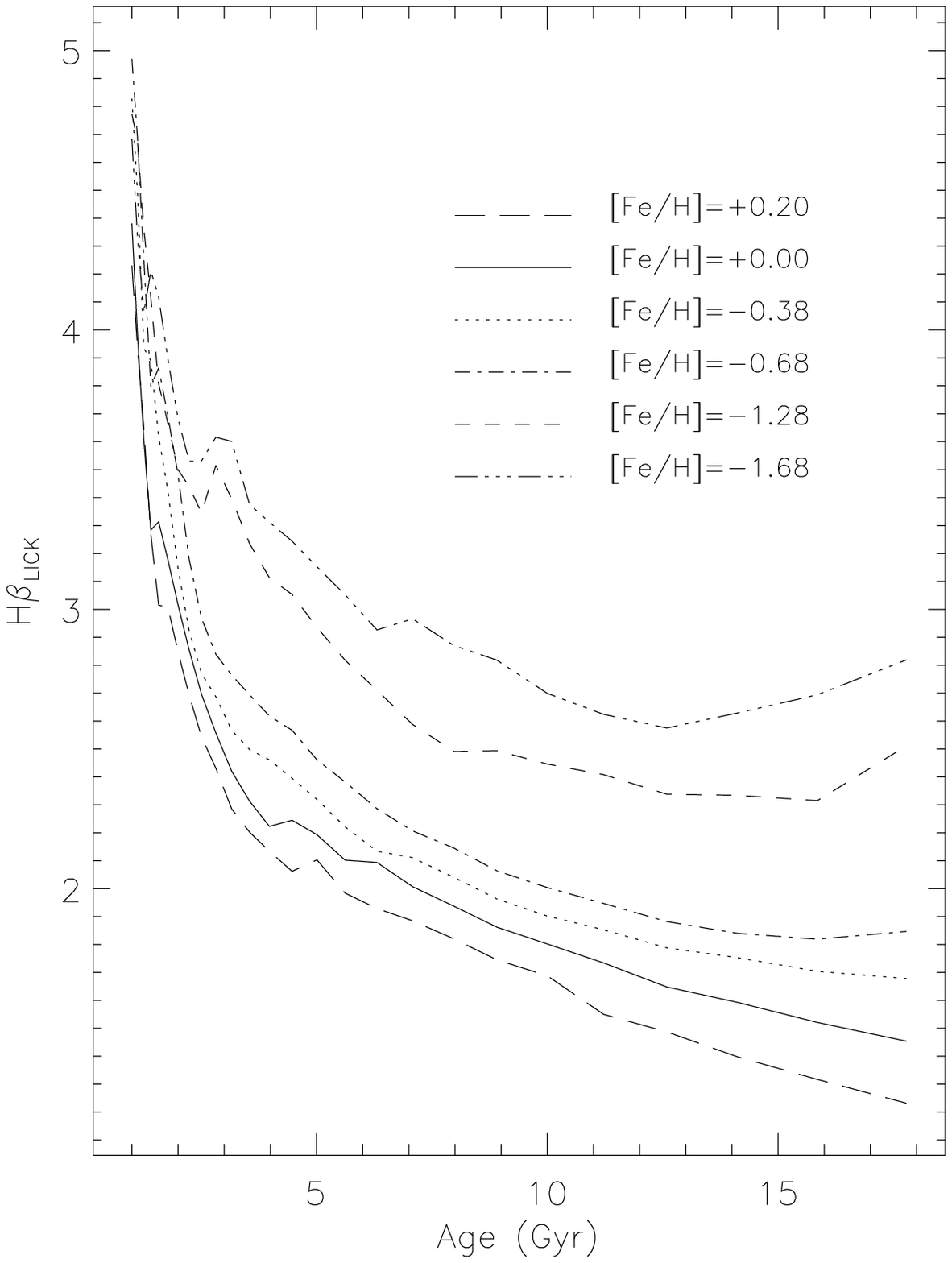}
\includegraphics[angle=0,scale=.50]{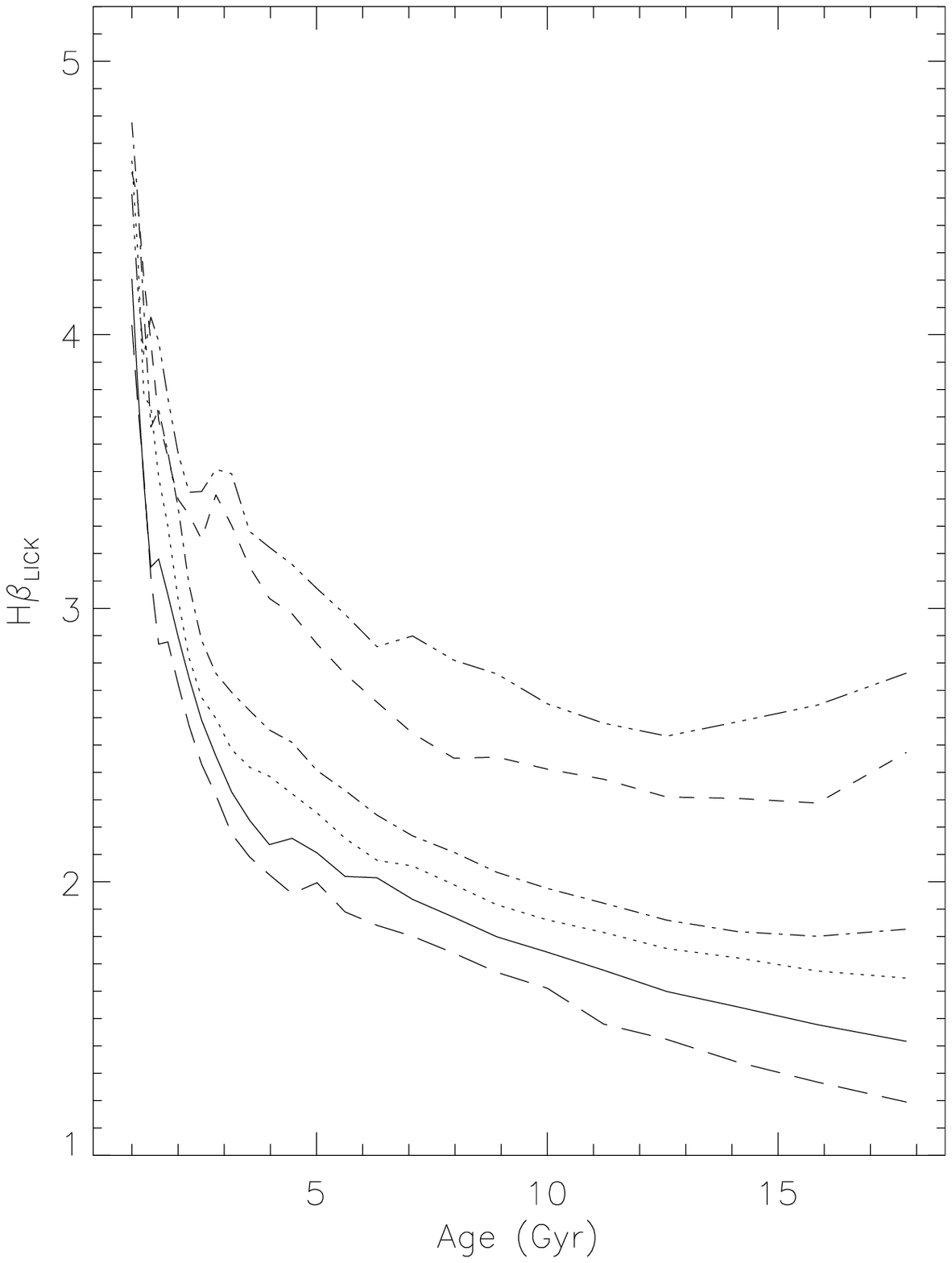}
\includegraphics[angle=0,scale=.50]{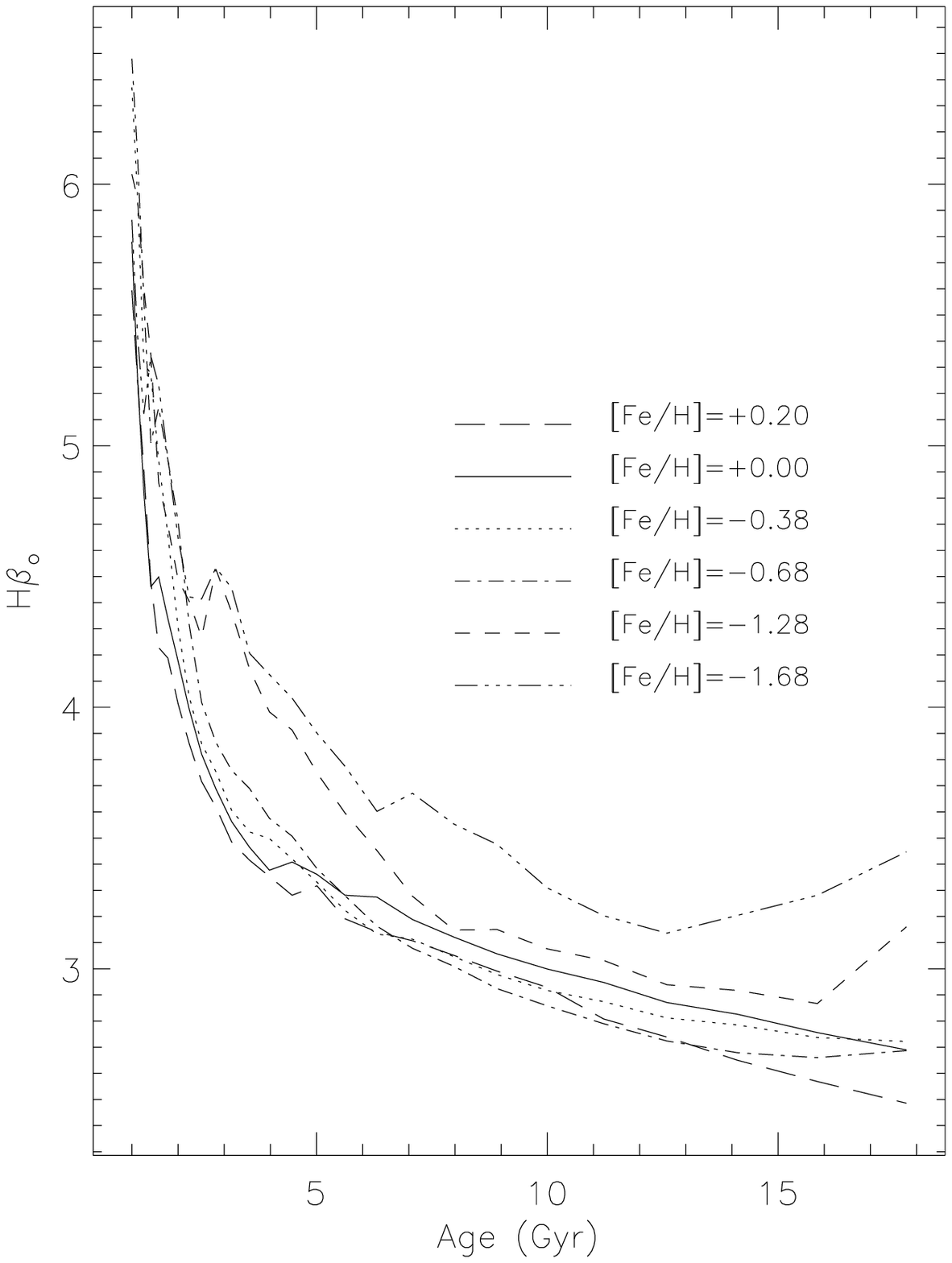}
\includegraphics[angle=0,scale=.50]{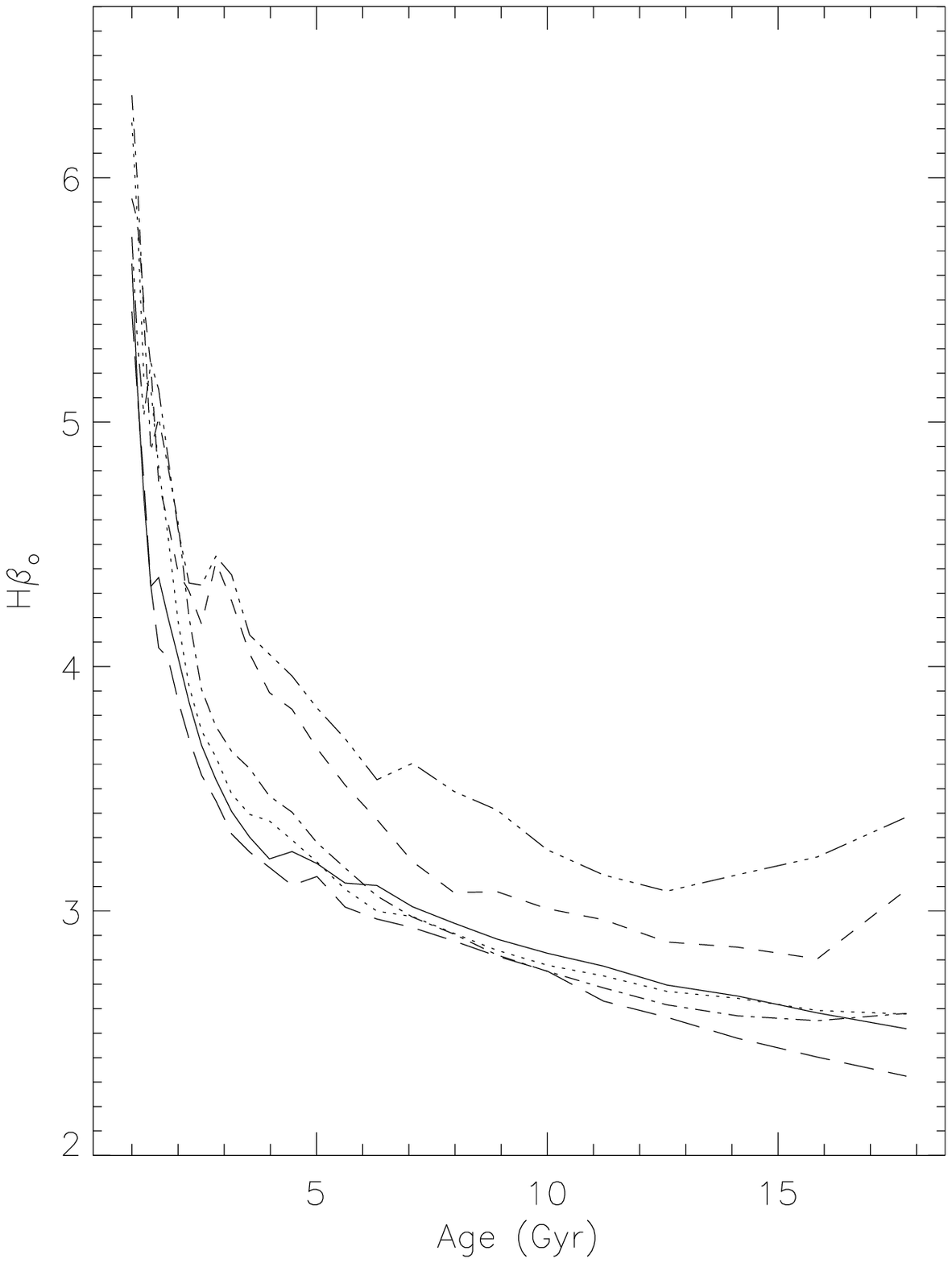}
\caption{H$_{\beta_{LICK}}$ and H$_{\beta_{o}}$ indices measured on the V07 SSP
spectral library at the nominal resolution of these models (left, FWHM=2.3\AA) 
and at a resolution similar to that of the Lick/IDS system (right, FWHM $\sim$
8.4\AA)}
\label{figure2} 
\end{figure*}

Figure \ref{figure2} shows H$_{\beta_{LICK}}$ and H$_{\beta_{o}}$ index values
as a function of the age of the SSP for different metallicities, {at the nominal resolution 
of the models (left) and at 225 \,Kms$^{-1}$ (right)\footnote{This broadening 
is similar to the Lick/IDS resolution at $\sim$ 5000 {\AA}} The plot shows
how H$_{\beta_{o}}$ is significatly less sensitive to the metallicity than
H$_{\beta_{LICK}}$ as the lines representing models of different metallicities lay
almost on the same curve, particularly for the higher metallicities (i.e. [M/H]$\geq$ -0.7)}. 
Note that this also applies for SSPs with ages below 5\,Gyr.
 
\subsection{Age-metallicity degeneracy}

We obtain for H$_{\beta_{o}}$ $\frac{\beta}{\alpha} < 0.01$. For comparison,
the standard H$_{\beta_{LICK}}$ index provides $\frac{\beta}{\alpha} \simeq
0.45$. (See Table \ref{table2}), confirming that the new index definition
increases significantly the age sensitivity. 

\begin{figure}
\includegraphics[angle=0]{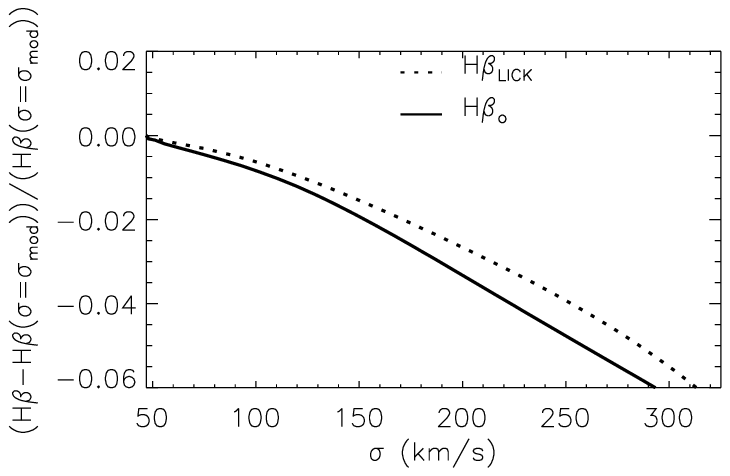}
\caption{Evolution of H$_{\beta_{LICK}}$ (dashed line) and  H$_{\beta_{o}}$
(solid line) versus spectral resolution ($\sigma$). The line represents  the
mean values for SSPs older than 1\,Gyr. Although H$_{\beta_{o}}$ index is more
dependent on $\sigma$, the variation is lower than 5\% at $\sigma <$ 250 ${\rm
km\,s^{-1}}$.} 
\label{figure3}
\end{figure}

\subsection{Spectral resolution and velocity dispersion}

H$_{\beta_{LICK}}$ presents a lower $\Sigma$ value than H$_{\beta_{o}} $(Table
\ref{table2} $\approx$ 0.1 against 0.19 for H$_{\beta_{o}}$). 

\begin{figure}
\includegraphics[angle=0]{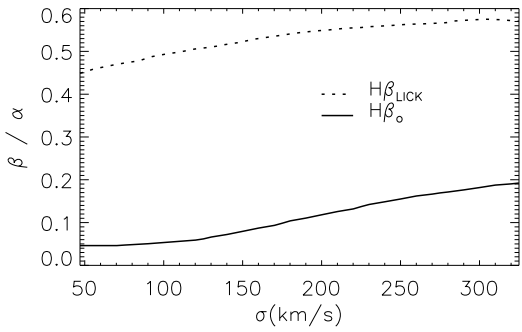}

\caption{Generalized Worthey's parameter, $\beta/\alpha$, versus model spectral
resolution ($\sigma$). This plot shows that H$_{\beta_{o}}$ is virtually insensitive to
metallicity as a function of $\sigma$ out to $\sigma \sim 300$. }

\label{figure4}
\end{figure}

Figure \ref{figure3} shows that both H$_{\beta_{LICK}}$ and H$_{\beta_{o}}$
depend very little on resolution. In fact this dependence is lower than 5 \%
when comparing the index value  at ${\rm 300\,km\,s^{-1}}$ and at the nominal
resolution of the models. Despite the fact of the slightly larger dependence of
H$_{\beta_{o}}$ index on $\sigma$, this dependence does not affect its age
sensitivity since $\frac{\beta}{\alpha}$ is always lower than 0.15, and much
lower than the value obtained for H$_{\beta_{LICK}}$ (See Figure
\ref{figure4}).

\subsection{Signal-to-Noise}

Table \ref{table2} lists $\sigma[I_{a}]$ in percentage. H$_{\beta_{o}}$ requires
slightly higher S/N than H$_{\beta_{LICK}}$ to distinguish the age once the
metallicity is known. 

\begin{figure}
\includegraphics[angle=0,scale=0.6]{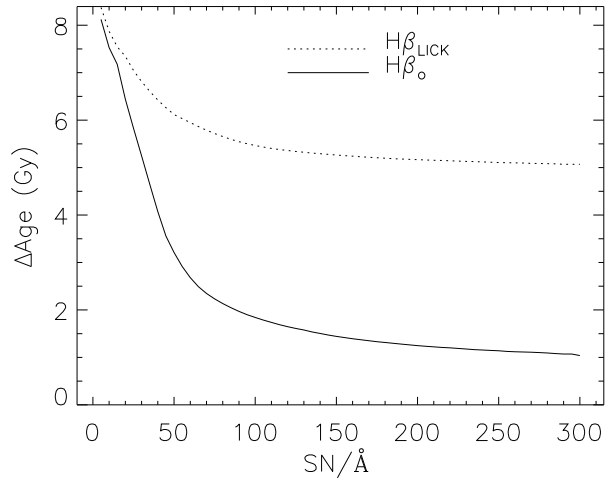}
\caption{Age accuracy determination for a 10\,Gyr solar metallicity SSP model as
a function of S/N (per {\AA}) obtained on the basis of the H$_{\beta_{LICK}}$
and H$_{\beta_{o}}$ indices. This plot shows that the maximum age accuracy
reached with H$_{\beta_{o}}$ is $\sim$ 1\,Gyr at SN/{\AA} $\sim$250, while an
age error of $\sim$3\, Gyr is obtained at SN/{\AA} $\sim$50. No matter the
spectrum quality, the maximum age accuracy for H$_{\beta_{LICK}}$ is $\sim$
5\,Gyr, if we neglect the metallicity information.}
\label{figure5}
\end{figure}

However, would we have considered the real error on the age,  its derivation
would have been different due to the fact that  any metallicity information had
been taken into account in $\sigma[I_{a}]$.  The real S/N requirements to derive
the age with precision  is tightly related to the metallicity dependence of the 
H$_{\beta}$ index definition. To estimate the age accuracy  determination for an
index definition when we neglect the metallicity,  we can focus for example on a
10\,Gyr model, which is  representative of the sub-space of ages older than
5\,Gyr and metallicities in the  range -0.7 $\leq$ [M/H] $\leq$ +0.2, where most
early-type  galaxies are located \citep{1998yCat..21160001T}. We then   simulate
the S/N effects on the spectrum and test the age  accuracy using a plot such as
that of Figure \ref{figure2}.  The obtained results are shown in Figure
\ref{figure5},  where the age uncertainity of H$_{\beta_{LICK}}$ decreases
asymptotically to $\sim$5.1\,Gyr with increasing S/N, while the minimun age
uncertainty associated to H$_{\beta_{o}}$ is $\sim$ 1\,Gyr at a S/N (per\,{\AA})
$\sim$250 \footnote{This plot can be used to prepare observations for galaxies
older than 5 \,Gyr and with metallicities higher than [M/H]=-0.7}. This means
that the maximun age accuracy achieved with H$_{\beta_{LICK}}$ is always lower
than that of H$_{\beta_{o}}$, no matter the spectrum quality, since this new
index is much less sensitive to metallicity.

\begin{table*}
 \centering
 \begin{minipage}{140mm}
  \caption{INDEX CHARACTERISTICS AND UNCERTAINTIES} 
  \label{table2}

  \begin{tabular}{@{}lrr@{}}
  \hline
{Parameter}    &  {H$_{\beta_{LICK}}$}   &
{H$_{\beta_{o}}$}  \\
  \hline
$\frac{\beta}{\alpha}$ & $\sim$ 0.45& $<$ 0.1\\
\\
$\Sigma$ ($10^{-4}$ ${\rm km\,s^{-1}}$)&1.05&1.89\\
Index variation by ${\rm \pm 10\,km\,s^{-1}}$ in $\sigma$ (\%) &  0.105 & 0.189 \\
Maximum stability $\sigma$ range& ${\rm > 300\,km\,s^{-1}}$ & ${\rm \pm 250\,km\,s^{-1}}$\\
\\
Age uncertainty caused by a $\Delta\lambda$ = 1.5 {\AA} &  5.5-18 \,Gyr&6.7-14.2
\,Gyr\\
Maximum $\Delta \lambda$ shifts ($\lambda$, z, rotation curve) compatible with a $+/-$ 2.5 \,Gyr error  ({\AA}) & 0.85&1.1\\
\\
(H$_{\beta}$(flux calibrated response curve) - H$_{\beta}$(continuum
removed))/H$_{\beta}$(f.c.r.c.) & $\lq$ 0.01 & $\lq$ 0.01 \\
\\
S/N (per\,{\AA}) required to distinguish 2.5 \,Gyr at 10 \,Gyr& $\sim$ 50  & $\sim$ 65\\ 
Minimum age uncertainity for S/N (per\,{\AA}) $>$ 250  & 5.1 \,Gyr & 1 \,Gyr \\ 
\hline

\end{tabular}
\end{minipage}
\end{table*}


\section{Characterization of H$_{\beta_{o}}$}

In this section we fully characterize H$_{\beta_{o}}$ index and discuss its
major uncertainties with several aspects that might influence its ability as an
age indicator.

\subsection{Sensitivity to abundance ratio variations}

An index definition includes the contribution of various chemical species,
despite the fact that a given element, which uses to name the index, might be
its major contributor. As giant elliptical galaxies show [Mg/Fe] overabundance
compared to the scaled-solar element partition we need to assess the influence
of such abundance ratios on our index definition and on its ability to
disentangle mean ages.

As this task is difficult to accomplish with models based on empirical stellar
spectra, the use of theoretical atmospheres to compute stellar spectra for a
large variety of element mixtures is an advantage \citep{1995AJ....110.3035T}.
In order to quantify the dependence of H$_{\beta_{o}}$ on [$\alpha$/Fe] we
should use SSP models with varying abundance ratios. We use the approach
described in \cite{2007IAUS..241..167C} as it is based on the same model that we
have employed here, where varying [$\alpha$/Fe] has been implemented via
differential correction making use of \cite{2005A&A...443..735C} stellar
library. Note that alternative SSP model SEDs computed with non scaled-solar
ratios have been recently published by \cite{2007MNRAS.382..498C}.  
The
\cite{2007IAUS..241..167C} model spectra cover two choices of partitions,
[$\alpha$/Fe] = 0.0 and 0.4, where O, Ne, Mg, Si, S, Ca and Ti are
flat-enhanced, whilst all the other elements follow Fe. The relation between the
total metallicity [M/H] and the iron content [Fe/H] is modified as\footnote{When
A = 0, we obtain the solar abundance of \cite{1998SSRv...85..161G}, adopted for
the synthetic spectra calculations of \cite{2005A&A...443..735C}}

\begin{equation}
\rm{[M/H]} = [Fe/H] + A[\alpha/Fe] 
\end{equation}

\noindent At a given [M/H], the sentitivity of an index to [$\alpha$/Fe] can
be evaluated as

\begin{equation}
{\Lambda}_{[\alpha/\rm{Fe}]} =
\Big[
\frac{|I_{[\alpha/{\rm Fe}]=0.4,[\rm{Fe/H}]=-0.3}-I_{[\alpha/{\rm Fe}]=0}|}{I_{[\alpha/{\rm Fe}]=0}}
\Big]_{[{\rm M/H}]=0;10\,Gyr}
\end{equation}

\noindent where $I_{[\alpha/{\rm Fe}]=0}$ is the index value for a 10\,Gyr solar
metallicity scaled-solar model, and $I_{[\alpha/{\rm Fe}]=0.4,[\rm{Fe/H}]=-0.3}$
is the index value for an $\alpha$-enhanced ([$\alpha$/Fe]=0.4) model of the
same age and total metallicity but [Fe/H]=-0.3. ${\Lambda}_{[\alpha/{\rm Fe}]}$
should tend to zero if an index does not depend on [$\alpha$/Fe].
${\Lambda}_{[\alpha/{\rm Fe}]}$ can be interpreted as a mean deviation caused by
the $\alpha$-enhancement with respect to the scaled-solar composition, which
translates to an age (or metallicity) uncertainty. Table \ref{table3} lists the
${\Lambda}_{[\alpha/{\rm Fe}]}$ values, and the associated age uncertainty,
corresponding to the various Balmer line index definitions ($\Lambda_{CO}$). We
find a greater sensitivity of H$_{\beta_{o}}$ to [$\alpha$/Fe] in comparison to
that of H$_{\beta_{LICK}}$. However this effect is smaller than that found for
the \cite{1997ApJS..111..377W} higher order Balmer index definitions.  

We are aware that the SSP SEDs with varying abundance ratios might depend on the
atmospheres and spectral synthesis codes employed to compute the theoretical
stellar spectra, which feed these models. In fact the computation of stellar
spectra with varying [$\alpha$/Fe] ratios requires adopting lists of all
relevant atomic and molecular transitions, along with accurate oscillator
strengths and damping constants. Furthermore fitting detailed line profile would
require the inclusion of NLTE, sphericity, chromosphere effects, among other
aspects. We therefore have extended the \cite{2007IAUS..241..167C} analysis by
including the $\alpha$-enhanced theoretical stellar library of
\cite{2005A&A...442.1127M} with the only purpose of assessing the uncertainties
affecting our result. As for \cite{2007IAUS..241..167C}, the model spectra cover
two mixtures, [$\alpha$/Fe] = 0.0 and 0.4, with O, Ne, Mg, Si, S, Ca and Ti
enhanced as in the \cite{2005A&A...443..735C} stellar library.  

Table \ref{table3} lists the results obtained for the ${\Lambda}_{[\alpha/{\rm
Fe}]}$ parameter based on this alternative library.  Surprisingly the two
libraries provide similar values for nearly all the Balmer line indices but
H$_{\beta}$. Unlike with the library of \cite{2005A&A...443..735C}, we obtain
that H$_{\beta_{o}}$ is less sensitive to $[\alpha/{\rm Fe}]$ variations than
H$_{\beta_{LICK}}$, (See Table \ref{table3},$\Lambda_{MU}$). Interestingly, the
H$_{\gamma_{\sigma}}$ index shows very little sensitivity to $[\alpha/Fe]$ for
the two libraries, which is not the case for the \cite{1997ApJS..111..377W}
indices. This result is in good agreement with \cite{2003MNRAS.339..897T}
conclusion for the latter indices. Such dependence of the H$_{\beta_{LICK}}$
index on [$\alpha$/Fe], when the library of \cite{2005A&A...442.1127M} is
employed, has been quoted before by \cite{2004MNRAS.353..917T}. This result is
however in disagreement with the abundance ratio insensitivity found by Tripicco
\& Bell (1995) and Korn et al. (2005) for H$_{\beta_{LICK}}$, as well as with
our own calculations on the basis of the library of \cite{2005A&A...443..735C}.
Therefore our conclusion on the greater sensitivity of H$_{\beta_{o}}$ on the
abundance ratio in comparison to H$_{\beta_{LICK}}$ must be taken with caveat,
as it might depend in part on the modelling of the stellar atmospheres. A
discussion of the feasibility of those stellar libraries is out of the scope of
this paper and we refer the reader to \cite{2007MNRAS.381.1329M} for an extended
analysis of these theoretical libraries and their use by SSP models. This
possible drawback of the new index definition is minimized by the fact that the
relative variation of H$_{\beta_{o}}$ to [$\alpha$/Fe] is very similar to that
of H$_{\beta_{LICK}}$ to the total metallicity. H$_{\beta_{o}}$ is however
completely safe for scaled-solar element partitions, with the advantage that it
provides almost orthogonal model grids. Finally, although not shown here, we
note that this orthogonality is preserved when [$\alpha$/Fe] enhanced models of
different ages and metallicities are employed to build-up the grids.

\subsection{Wavelength and radial velocity uncertainties} 

\begin{figure}
\includegraphics[angle=0,scale=0.5]{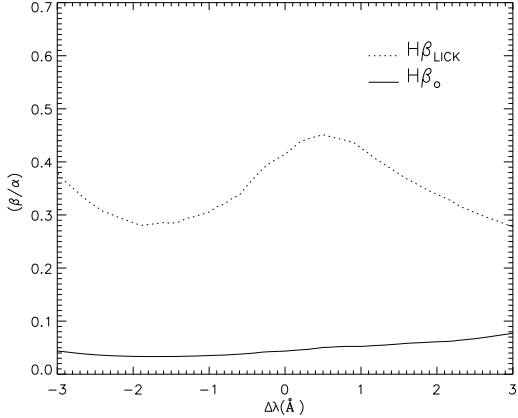}
\caption{Generalized Worthey's parameter, $\beta/\alpha$, versus wavelength
shifts. This plot shows that the age sensitivity of H$_{\beta_{o}}$ depends very
little on this parameter.}
\label{figure11}
\end{figure}

As absorption line indices provide us with information from relatively narrow
bandpasses, errors in wavelength calibration, radial velocity or rotation
curve, lead to errors in our age/metallicity estimates. It is worth to note
that an accurate wavelength calibration is tipically around 5\% - 10\% of the
dispersion resulting from the adopted instrumental setup employed in the
observations. To account for this we obtain the largest wavelength errors or
shifts allowed to achieve a minimum precision of 2.5\,Gyr on age for a solar
metallicity SSP model of 10\,Gyr. In principle we should have discarded index
definitions that are unstable for wavelengths shifts of 0.5\,{\AA} (i.e.
tipically corresponding to  dispersion of 5\,{\AA}), according to our
criterion. However the set of preselected definitions that were virtually
independent on metallicity did not show such sensitivity to wavelength shifts.
For this reason we only characterize the new H$_{\beta_{o}}$ index for
wavelength shifts and compare the obtained result with that for
H$_{\beta_{LICK}}$. Table \ref{table2} lists the largest wavelength errors
allowed to obtain a minimum precision of 2.5\,Gyr for a solar metallicity,
10\,Gyr old SSP model for these two indices. This table also lists the age
uncertainty corresponding to a wavelength shift of 1.5\,{\AA} (i.e.
${\rm \sim 90\,km\,s^{-1}}$). Figure \ref{figure11} shows how the $\frac{\beta}{\alpha}$ parameter
changes with increasing wavelength shift. We see that wavelength shifts do not
affect  significantly the age resolving power of these two indices. Note that
the age-sensitivity of H$_{\beta_{o}}$ is always larger than that of
H$_{\beta_{LICK}}$. Interestingly, H$_{\beta_{LICK}}$ increases its ability for
disentangling ages if its bandpasses are shifted 1.7\,\AA\ blueward. The main
characteristics and capabilities of this alternative H$_{\beta_{LICK}}$ index
definition are sumarized in the Appendix.

\subsection{The Spectrum Shape} 

We have tested the effect of the spectral response curve on the H$_{\beta}$
indices. We expect this effect to be more significant for H$_{\beta_{o}}$ in
comparison to H$_{\beta_{LICK}}$. This is because the definition of
H$_{\beta_{o}}$spans a $\sim$ 30 {\AA} wider spectral range. We follow the the
test proposed by \cite{1986A&A...164..260A} and show in Table \ref{table2} the largest
differences, obtained for the oldest SSPs, when comparing the index mesurements
performed on the flux calibrated and continuum removed SSP spectra (using a
spline3 of order 4) of similar age and metallicity. We conclude that the effect
of the continuum shape on the index measurements is negligible for these two
indices.  

\subsection{Dust extinction} 

Recently, absorption-line studies of integrated stellar populations are being
extended to later type galaxies which may contain significant amounts of dust
(e.g., \cite{2007A&A...474.1081G,2007IAUS..241..420D}).
In the case of H$_{\beta_{LICK}}$, when dust extinction affects the SSP age
determination, the errors on the physical parameters are of the same order as
the ones measured in the index and thus would not likely be detected above the
noise \citep{2005ApJ...623..795M}. We should not expect significant differences
for H$_{\beta_{o}}$, though we have performed the same analysis followed by
\cite{2005ApJ...623..795M} who takes into account the two-component model of
\cite{2000ApJ...539..718C} for the influence of the interstellar medium on the 
starlight. The two adjustable parameters of this model are  $\tau_{V}$, the
total effective V-band optical depth affecting stars younger than $10^{7}$ \,yr 
and $\mu$, the fraction of the total dust absorption contributed by diffuse
interstellar medium dust\footnote{See \cite{2005ApJ...623..795M} for a full
description of the method}. 

\begin{figure*}
\center
\includegraphics[angle=0,scale=0.6]{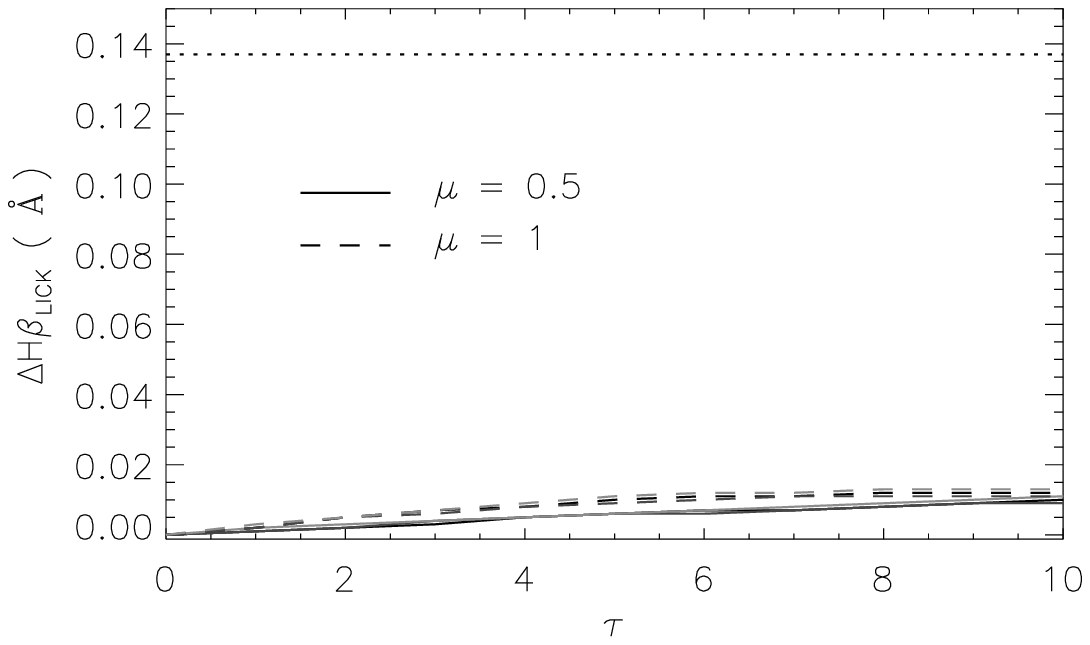}
\includegraphics[angle=0,scale=0.6]{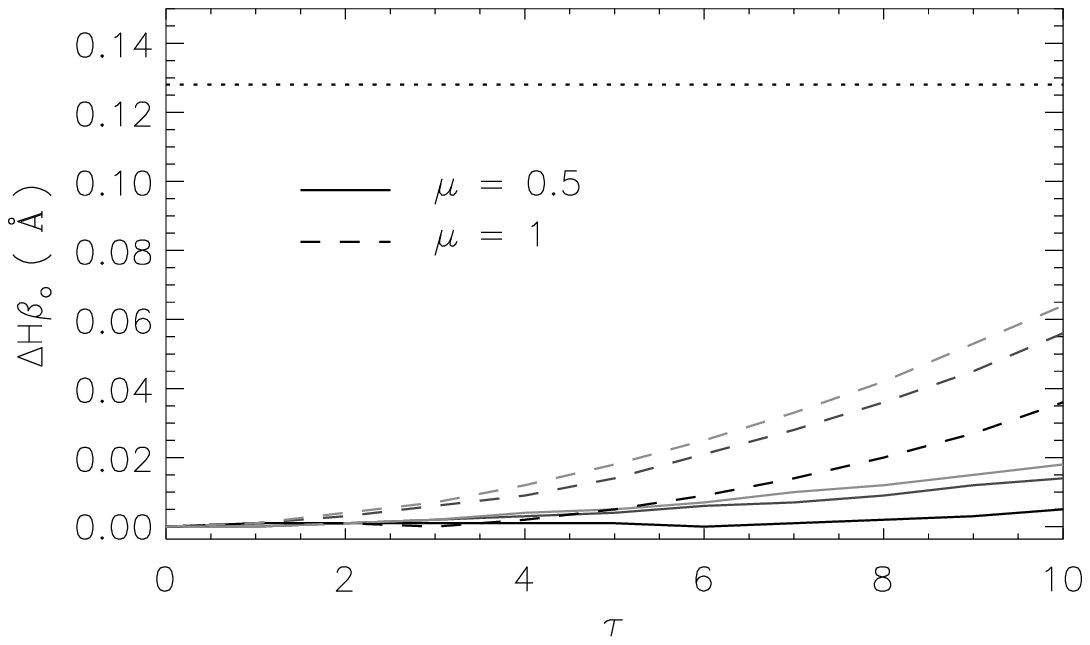}
\caption{H$_{\beta_{LICK}}$ and  H$_{\beta_{o}}$ index residuals, $\Delta index
= index(\tau_{V}) - index(\tau_{V} = 0)$, as a function of the total effective
V-band optical depth, $\tau_{V}$, for solar metallicity SSPs. The results for
models of 1 (light grey lines), 5 (grey lines), and 13\,Gyr (black lines) are
shown. Different values of  $\mu$ are represented as 0.5 (solid line), and 1.0
(dashed line).  The black horizontal dotted lines represent the errors required
for distinguishing 2.5\,Gyr for a solar metallicity SSP model of 10\,Gyr.}
\label{figure6}
\end{figure*}

The variations of H$_{\beta_{o}}$ as a function of $\tau_{V}$, $\mu$ and age
for a solar metallicity SSP model is shown in Figure \ref{figure6}, where
$\Delta index$ versus $\tau_{V}$ is plotted. In this figure $\Delta index$ is the
difference between the index measured with and without dust, i.e.,$\Delta
index = index(\tau_{V}) - index(\tau_{V} = 0)$. Results are shown for model
ages of 1 (light grey lines), 5 (grey lines) and 13\,Gyr (black line). The two
values for $\mu$ of 0.5 and 1.0 correspond to the solid and dashed lines,
respectively. The black horizontal dotted line represents the measurement error
required to distinguish 2.5\,Gyr for a 10\,Gyr solar metallicity SSP model, 
i.e., the accuracy we have imposed during our process of finding the new 
H$_{\beta_{o}}$ index. For a 1\,Gyr scaled-solar 
model (light grey lines), H$_{\beta_{LICK}}$ $\sim$ 4.3 \,{\AA} and 
H$_{\beta_{o}}$ $\sim$ 5.5\,{\AA}, it corresponds to 1\% for H$_{\beta_{o}}$ 
and 0.2\% for H$_{\beta_{LICK}}$, in the two cases for a extreme dust extintion. 
For a 13\,Gyr model (black lines), H$_{\beta_{LICK}}$ $\sim$ 1.8\,{\AA} 
and H$_{\beta_{o}}$ $\sim$ 3 \,{\AA}, it means 0.3\% for H$_{\beta_{o}}$, 
since $\Delta index$ $\sim$ 0.01 \,{\AA} and 0.6\% for H$_{\beta_{LICK}}$.

As expected, the effect of extinction is larger in H$_{\beta_{o}}$ than in
H$_{\beta_{LICK}}$, since the involved wavelength coverage is larger. However,
the obtained effects are lower than the minimum index errors associated to the
noise (Figure \ref{figure6}), when extinction has been incorporated on top of
the SSP SEDs. It is worth noting than although dust extintion was not 
considered as a parameter for the index definition proccess, it is not rejecting
a  posteriori the obtained index.

\section{Discussion}
 
Estimating the mean luminosity-weighted age of an early-type galaxy represents
a major step for constraining its Star Formation History. In this section we
probe the  H$_{\beta_{o}}$ as an age-dating indicator in real data. We apply
this index to Milky Way globular clusters and to prototype early-type galaxies
for which extremely high quality spectra are available. 

\subsection{Galactic Stellar Clusters}

Globular clusters are ideal laboratories to test the SSP models as they can be
considered as stellar populations formed in a single and homogenous process. 
Milky Way globular clusters allow us to check the consistency of the
H$_{\beta_{o}}$ age estimates as it is possible to obtain independent
age/metallicity values from detailed Color-Magnitud Diagram (CMD) analyses.

\begin{figure*}
\center
\includegraphics[angle=0,scale=0.9]{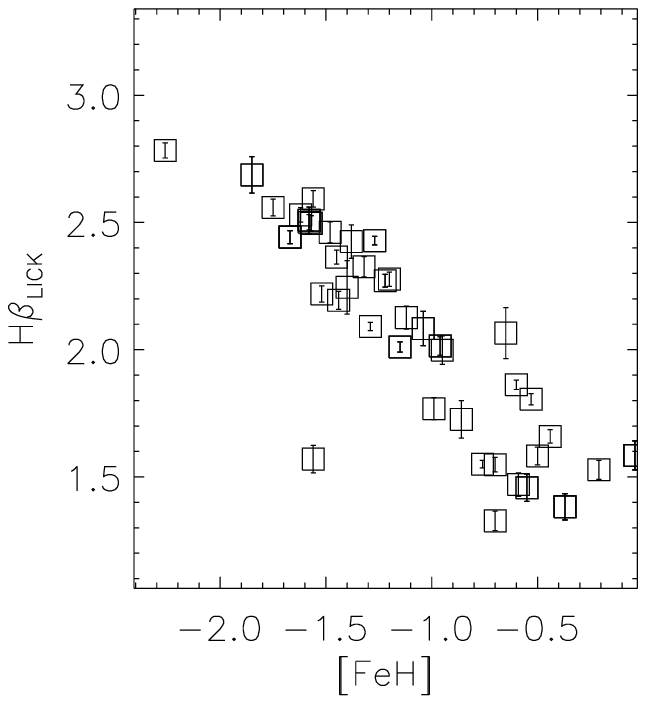}
\includegraphics[angle=0,scale=0.9]{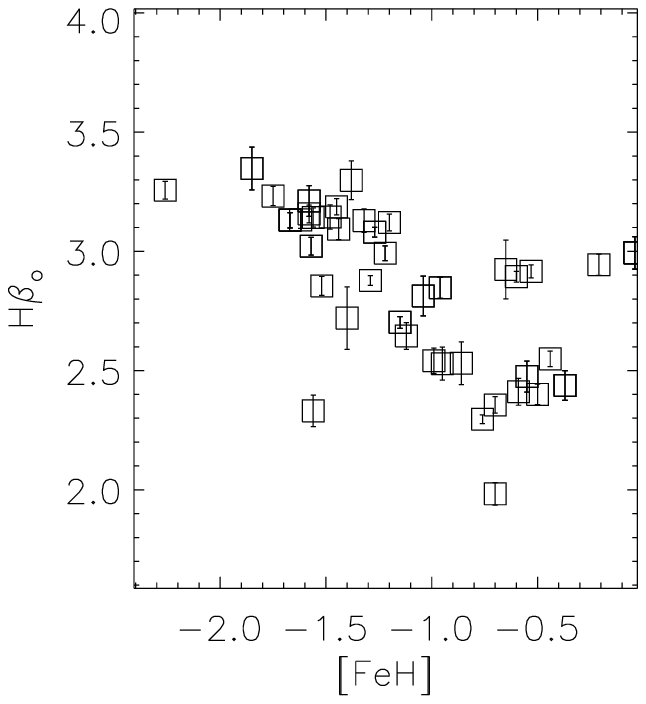}
\caption{H$_{\beta_{LICK}}$ (left) and H$_{\beta_o}$ (right) vs. CMD-derived 
metallicities for the globular cluster sample of Schiavon et al. (2005).
The [Fe/H] values were taken from the compilation of Harris 1996.} 
\label{figure10}
\end{figure*}

Figure \ref{figure10} shows the H$_{\beta_{LICK}}$ and H$_{\beta_{o}}$ indices
measured on the Milky Way cluster sample of  \cite{2005ApJS..160..163S} versus
the CMD-derived metallicities. The [Fe/H] values are taken from
\cite{1996AJ....112.1487H} (the 2003 version of the McMaster catalog). We
obtain for the H$_{\beta_{LICK}}$ sequence a Spearman rank coefficient value of
-0.89, whereas for H$_{\beta_{o}}$ we obtain -0.70, indicating a milder
anti-correlation for the latter. The slope of a linear fitting is -0.71
\,dex/{\AA} for H$_{\beta_{LICK}}$ and -0.36 \,dex/{\AA} for H$_{\beta_{o}}$. Note
that the fit for H$_{\beta_{o}}$ is virtually flat for [Fe/H]$>$-1.0. 
Although we have shown in the previous section that H$_{\beta_{o}}$ 
is particularly optimized for higher metallicities ([M/H]$\geq$-0.7),  
these results confirm the lower metallicity dependence of H$_{\beta_{o}}$ with
respect to H$_{\beta_{LICK}}$ on the basis of real data, without the use of
models, for all metallicities.

\begin{figure}
\includegraphics[angle=0,scale=0.85]{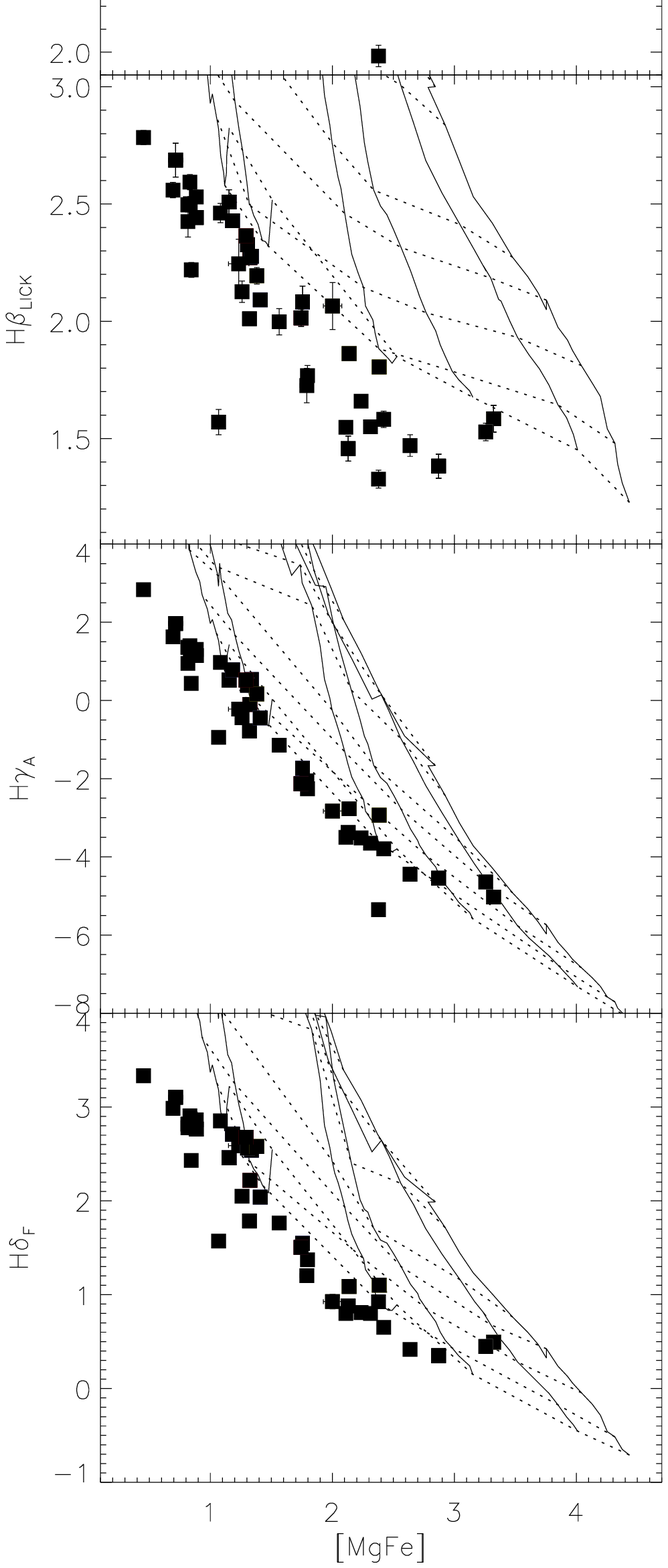}
\caption{From top to bottom H$_{\beta_{o}}$, H$_{\beta_{LICK}}$,
H$_{\gamma_{A}}$  and H$_{\delta_{F}}$ versus [MgFe] index for the
Schiavon et al. (2005) sample of GCs (top to bottom). Model grids smoothed
to the resolution of the instrumental setup employed in the data are
overplotted. Metallicity increases from left to right (solid lines): [Fe/H] =
-1.7, -1.3, -0.7, -0.4, 0.0 and +0.2. 
Age increases from top to bottom (dotted lines): Age = 3, 5, 8, 13 and 18\,Gyr.} 
\label{figure7}
\end{figure}

Figure \ref{figure7} shows H$_{\beta_{o}}$ and three age-dating indices versus
the mean metallicity indicator [MgFe] for the same cluster sample.   We see that
the stellar clusters fall at the bottom of all these plots, indicating very old
ages. In fact for many cases we obtain ages that are larger than the oldest
models (i.e. older than the age of the Universe). These plots show the well
known model zero-point problem affecting the spectroscopic age determinations
from the Balmer line indices (e.g. \cite{1999AJ....118.1268G,
2001ApJ...549..274V,  2002ApJ...580..850S}).  Recently,
\cite{2007MNRAS.379.1618M} using multi-index  $\chi^{2}$ minimization technique
analized two Milky Way globular cluster samples (the Schiavon's data used here
and Puzia et al. 2002) to test the age estimates obtained with three different
sets of updated models  (Thomas et al. 2004, Lee \& Worthey 2005 and our
models).  Although the ages inferred on the basis of their multi-index method
are  in agreement with the CMD-derived ages, their Figure 2 shows basically the
same result, i.e. the observed H$_{\beta}$ values fall below the model grids for
the three models.  Nonetheless the large age values is a result that is common
to all the Balmer line-strength index definitions shown in Figure \ref{figure7}.
However there are other works, working in the Lick/IDS system, in which this
offset is not seen, e.g. Thomas et al. (2003) where these models are used to fit
the Puzia et al.'s data (both, the models and the data, also employed in
\cite{2007MNRAS.379.1618M}). Note also that according to the $\Lambda_{CO}$
value listed in Table \ref{table3} for H$_{\beta_{o}}$ this offset is clearly
minimized when employing models based on the \cite{2005A&A...443..735C} library.
As this problem does not affect the relative age/metallicity sensitivities of
the H$_{\beta}$ index definitions used here we refer the interested reader to
the above papers for further details on this issue. 

In the H$_{\beta_{o}}$ plot we are able to distinguish two populations of
clusters for [MgFe] $>$ 1.5 {\AA}, which can also be seen in Figure
\ref{figure10}. With this evidence in our hands we can identify these two
cluster populations in the H$_{\beta_{LICK}}$ plot as well. However the two
cluster populations cannot be distinguished in the remaining two panels of
Figure \ref{figure7}, which include lower age-sensitivity indices. Therefore
this intriguing feature becomes more evident as the metallicity dependance of
the Balmer index definition has been significantly minimized. An extensive
study devoted to understand the origin of this feature is presented in
\cite{cenarrosb}. In this study we compare the CMDs of these
clusters and discuss among other aspects, the effects of the Horizontal Branch
morphology and the Blue Stragglers on the integrated H$_{\beta}$ indices. In
that paper we convincingly show that the latter is directly linked to the
observed two age-population feature and that this feature is not driven by the
zero-point problem.   

\begin{table*}
 \centering
 \begin{minipage}{140mm}
  \caption{INDEX SENSITIVITY TO [$\alpha$/Fe] VARIATIONS}
  \label{table3}
  \begin{tabular}{@{}lrrrrr@{}}
  \hline

Index   &  $\Lambda_{CO}$ &
Age uncertainity & $\Lambda_{MU}$ &
Age uncertainity\\
   &   &
at 10 \,Gyr (in \,Gyr) & &  at 10 \,Gyr (in \,Gyr)  \\

  \hline
H$_{\beta_{LICK}}$& 0.011  & $<$ 1 &0.079 & 8-13\\
H$_{\beta_{o}}$& 0.119  & 5-18&$<$0.01& $<$ 1 \\
H$_{\sigma_{130}}$& 0.045 &  8-12 &0.020& 9-12\\
H$_{\delta_{A}}$& 0.75  & 2.5-18 &1.1 & 1-18\\
H$_{\gamma_{A}}$& 0.23  & 4-18 &0.17& 3.5-18\\
H$_{\delta_{F}}$& 1.7  &3.5-18 & 0.4& 7-15\\
H$_{\gamma_{F}}$& 0.6 &3.5-18 & 0.5& 3.5-18 \\
  
  \hline
\end{tabular}
\end{minipage}
\end{table*}

\subsection{Elliptical galaxies}

In this section we use the H$_{\beta_{o}}$ index to estimate the ages of a well
selected sample of ellitpical galaxies. For selecting the sample we mainly
followed two criteria: a wide coverage in galaxy mass and availability of
spectra of very high S/N.  The galaxy sample is composed of M\,32
\citep{1985AJ.....90.1927R} and six elliptical galaxies of Virgo, selected
along the Color-Magnitude Relation of this cluster \citep{2001ApJ...551L.127V}.
The long-slit spectra for all the galaxies have S/N (per \AA) above 150.  In
Figures 8 and \ref{figure9}, we show H$_{\beta_{LICK}}$ and H$_{\beta_{o}}$ as
a function of various metallicity indicators: [MgFe], Fe$_{3}$, Mg$_{\rm{b}}$,
Ca4227 and CN$_{2}$. Galaxies with similar velocity dispersions were grouped
separately:  $\sigma_{group}$ $\approx$ 135, 180, 225  km\,s$^{-1}$. To allow
direct comparations between the galaxies having similar velocity dispersions,
some small $\sigma$ corrections were applied:  the galaxies with
$\sigma_{total}[=(\sigma^{2}_{galaxy}+\sigma^{2}_{instr.})^{1/2}] <
\sigma_{group}$ were convolved with the appropiate Gaussian to reach the
corresponding $\sigma_{group}$. 

\begin{figure*}
\includegraphics[angle=0]{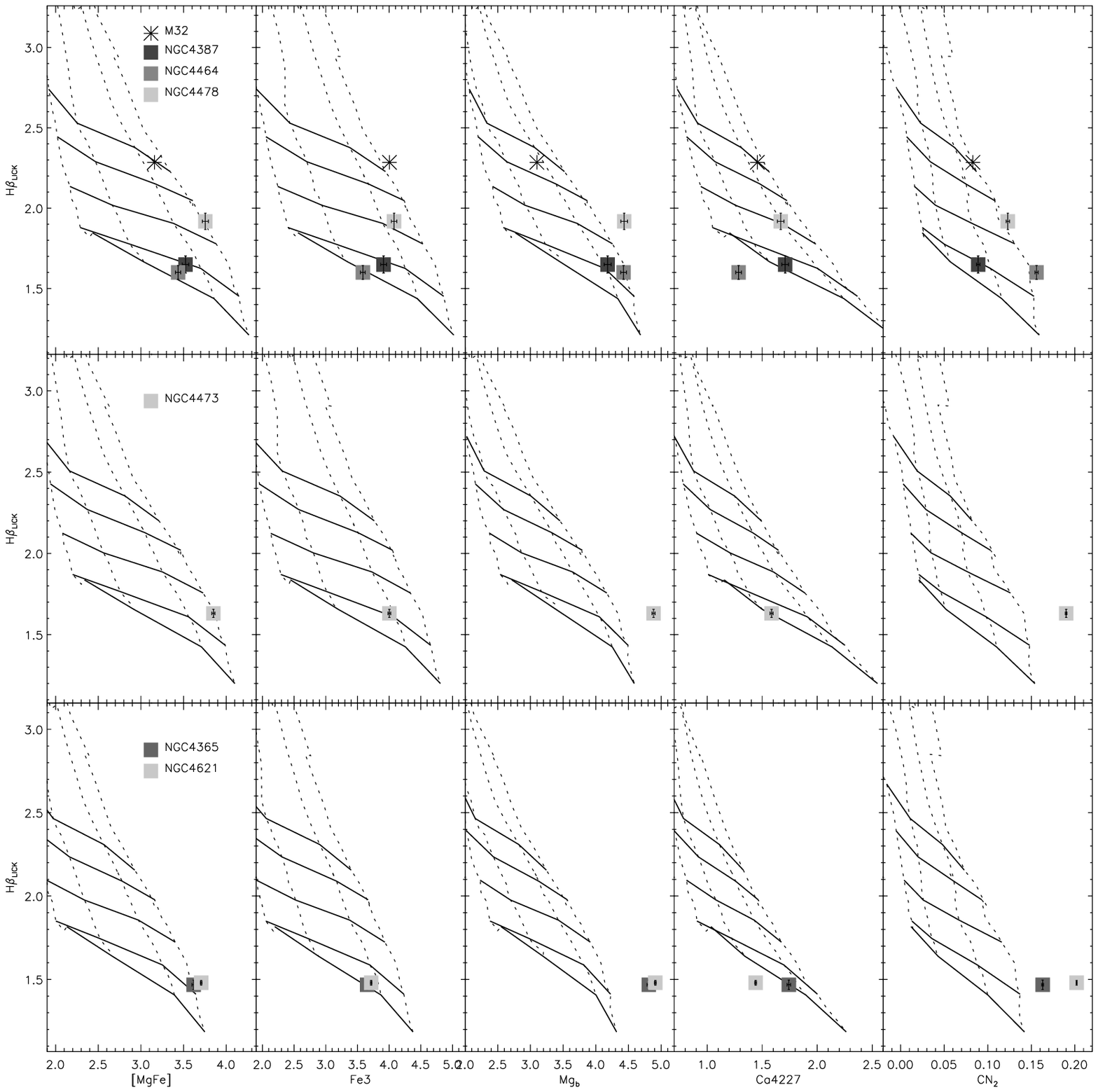}
\caption{[Mg/Fe], Fe$_{3}$, Mg$_{b}$, Ca4227 and CN2  vs. H$_{\beta_{LICK}}$.
Top, middle and bottom panels are for galaxies with $\sigma_{group}$ = 135, 180,
and 300 ${\rm km\,s^{-1}}$, respectively. Model grids with various ages (dotted
lines) and metallicities (solid lines) are overplotted; the age increases from
top to bottom (5.6, 8.0, 11.2 and 17.8 \,Gyr), whereas the metallicity increases
from left to right ([Fe/H]=-0.7, -0.4, 0.0 and +0.2).}
\label{figure8}
\end{figure*}

\begin{figure*}
\includegraphics[angle=0]{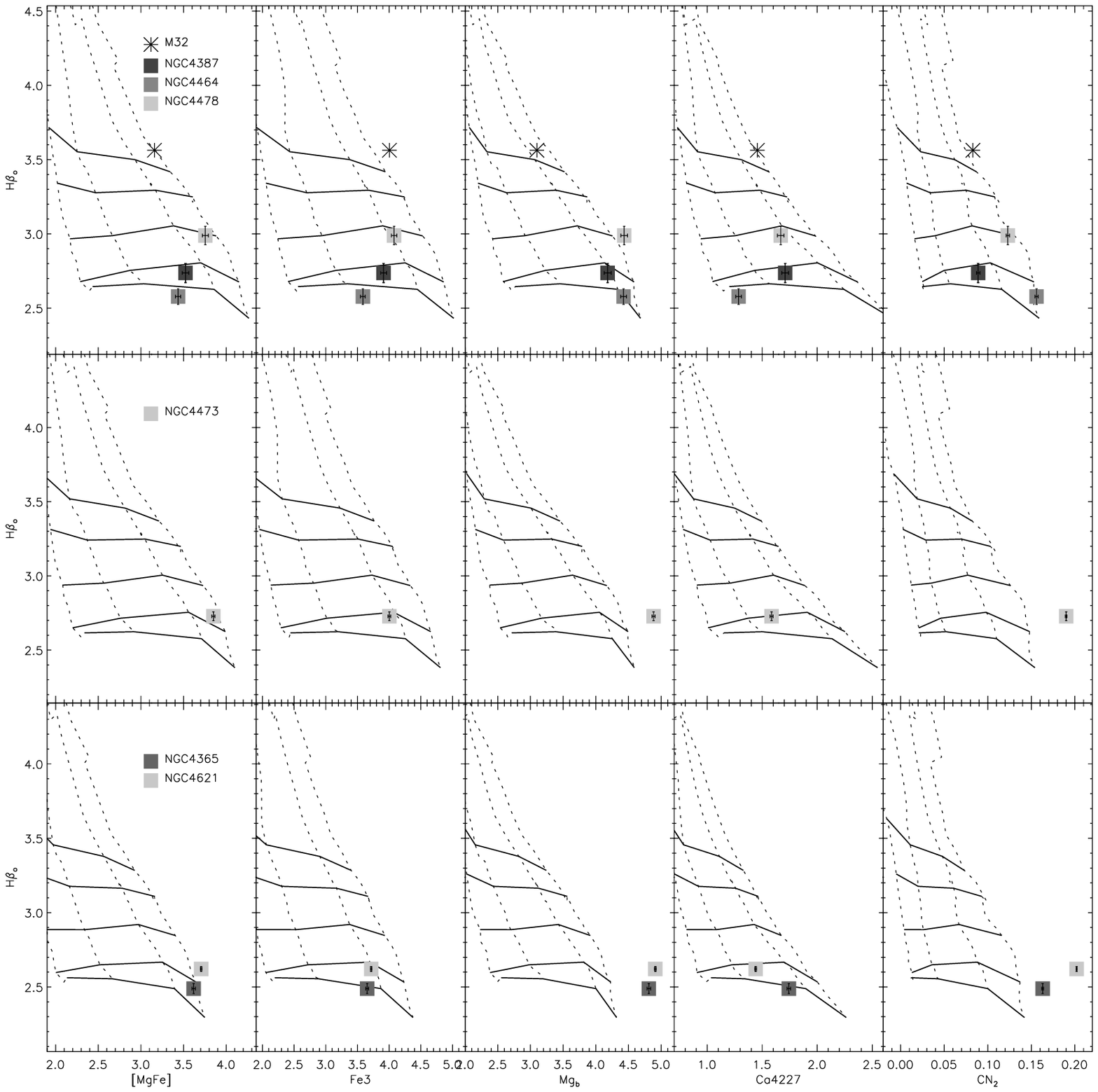}
\caption{Same as Figure \label{fig8} but for H$_{\beta_{o}}$.}
\label{figure9}
\end{figure*}

It is commonly used the H$_{\beta_{LICK}}$ versus [MgFe] to determine both the
age and the total metallicity,  as the latter has been shown to be rather
insensitive to possible non scaled solar element ratios which are found in
massive ellipticals \citep{2004MNRAS.351L..19T}. However as H$_{\beta_{LICK}}$
has some sensitivity to metallicity the age estimates depend on the metallicity
indicator in use, i.e. younger for a Magnesium dominated index and older for an
Iron dominated index. This problem is usually alleviated through an iterative
process with the aid of models that specifically take into account the non
scaled-solar ratios (e.g. \cite{2004MNRAS.353..917T, 1998yCat..21160001T,
2003MNRAS.339..897T}).  This is no longer a problem if we use H$_{\beta_{o}}$
index as it can be seen in Figure \ref{figure9}. This figure shows that the
derived ages are consistent irrespectivly of the metallicity indicator in use. 
Table \ref{table2} lists the  H$_{\beta_{LICK}}$ and H$_{\beta_{o}}$ age
estimates for M\,32 and the Virgo galaxies as derived from all the diagramas in
\ref{figure8} and \ref{figure9}.  

Our H$_{\beta_{o}}$ age estimates are consistent within the error bars with
those derived by \cite{1985AJ.....90.1927R} for M\,32 and
\cite{2001ApJ...551L.127V} and \cite{2006ApJ...637..200Y} for the Virgo sample
using H$_{\gamma_{\sigma}}$ high S/N requirement set of indices.  This sample
of high quality spectra confirms H$_{\beta_{o}}$ as an advantageous age-dating
indicator. A more detailed analysis of the ages and metallicities of these
galaxies has already been performed by these authors.

\section{Conclusions}

We have defined a new spectroscopic age indicator, H$_{\beta_{o}}$, which has
been optimized for disentangling stellar cluster and galaxy ages. This index
has a larger ability for lifting the age-metallicity degeneracy than the
standard index H$_{\beta_{LICK}}$. To achieve this, we have employed the
evolutionary stellar population synthesis model of \cite{1999ApJ...513..224V}
and its recent extension \cite{vazdekis} based on MILES stellar spectral
library (S\'anchez-Bl\'azquez et al. 2006; Cenarro et al. 2007). As these
models provide full spectra at moderately high resolution for stellar
populations of different ages and metallicities, it is straighforward to
investigate the behaviour of prospective index definitions as a function of
relevant parameters by measuring the indices directly on the SSP spectra. This
avoids us going through an intermediate step that requires the parametrization
of each index definition as a function of the stellar atmospheric parameters to
compute the integrated index, as it has been done, for example, for the
Lick/IDS system of indices. The latter approach is not functional for
optimizing trial index definitions as a function of age or metallicity, or
other effects such as velocity dispersion. 

We have shown that the stronger age disentangling power of H$_{\beta_{o}}$ is
achieved by avoiding the metallic lines of the red-pseudocontinuum of
H$_{\beta_{LICK}}$. The main characteristics and uncertainties affecting
H$_{\beta_{LICK}}$ and H$_{\beta_{o}}$ indices have been studied in detail. We
find that H$_{\beta_{o}}$ has a slightly higher velocity dispersion sensitivity
than H$_{\beta_{LICK}}$, but this effect is negligible in comparison to that
from photon noise. The S/N required to measure H$_{\beta_{o}}$ is not
significantly higher than that for H$_{\beta_{LICK}}$, and much lower than that
needed for applying H$\gamma_{\sigma}$ set of age indicators of
\cite{1999ApJ...525..144V}.  We have fully characterized the behaviour of
H$_{\beta_{o}}$ with wavelengths shifts, spectrum shape and dust extinction. We
find that none of these effects are particularly relevant.  We also have studied
the effects of the [$\alpha$/Fe] enhancement on H$_{\beta_{o}}$ making use of
the theoretical stellar spectral libraries of \cite{2005A&A...443..735C} and
\cite{2005A&A...442.1127M} following the approach of \cite{2007IAUS..241..167C}.
We find a greater sensitivity of H$_{\beta_{o}}$ to [$\alpha$/Fe] in
comparison to that obtained for H$_{\beta_{LICK}}$ when the library of
\cite{2005A&A...443..735C} is employed. Furthermore the relative variation of
H$_{\beta_{o}}$ to [$\alpha$/Fe] is very similar to that of H$_{\beta_{LICK}}$
to the total metallicity, but this result must be taken with caution as the
opposite trend is obtained when employing an alternative stellar library
(\cite{2005A&A...442.1127M}). The sensitivities of H$_{\beta_{o}}$ and
H$_{\beta_{LICK}}$ to  [$\alpha$/Fe]  variations are smaller than those found
for the \cite{1997ApJS..111..377W} higher order Balmer index definitions.
Interestingly the H$_{\gamma_{\sigma}}$ indices of \cite{1999ApJ...525..144V}
and \cite{2001ApJ...549..274V} show negligible sensitivity to [$\alpha$/Fe].

We have analyzed the Milky Way globular cluster spectra of
\cite{2005ApJS..160..163S} to test H$_{\beta_{o}}$. The plots of their
CMD-derived metallicities against H$_{\beta_{LICK}}$ and H$_{\beta_{o}}$
indices, show how the metallicity sensitivity has been decreased significantly
for latter. The comparison of the observed values and the model grids resulting
from plotting H$_{\beta_{o}}$ versus the [MgFe] metallicity indicator provide
very old ages, in good agreement with the results obtained from various Balmer
age indicators, when employing scaled solar SSP models. However the obtained
ages are older than the CMD-derived ages, confirming a model zero-point problem
that might be affecting the analyses of the integrated light based on the Balmer
indices (e.g. \cite{1999AJ....118.1268G, 2001ApJ...549..274V, 
2002ApJ...580..850S, 2007MNRAS.379.1618M}).

We also probe the reliability of H$_{\beta_{o}}$ to obtain mean luminosity
weighted ages of early-type galaxies. For this purpose we used a sample of
ellipticals covering a wide range in mass and $\alpha$-enhancement, for which
spectra of extremely high quality are available
\citep{2005AJ....129..712R,2001ApJ...551L.127V}. Unlike with H$_{\beta_{LICK}}$,
the ages inferred from plotting H$_{\beta_{o}}$ versus various metallicity
indicators and scaled-solar model grids are consistent irrespective of the
metallicity indicator in use. This also applies to the more massive galaxies
with larger [Mg/Fe] values. We also find that the H$_{\beta_{o}}$ ages are in
good agreement with the values obtained from the very high S/N requirement
H$\gamma_{\sigma}$ set of indices of \cite{1999ApJ...525..144V} as listed in
\cite{2006ApJ...637..200Y} on the basis of the same SSP models.

The results and plots shown here might be very usefull, and could be taken as a
guide, for preparing and optimizing observations for those willing to use
H$_{\beta_{o}}$ and H$_{\beta_{LICK}}$ indices in their analyses.

\begin{table*}
 \centering
 \begin{minipage}{140mm}
  \caption{AGE ESTIMATES FOR VIRGO ELLIPTICAL GALAXIES} \label{table5}
  \begin{tabular}{@{}rrrrrrrrrrrr@{}}
  \hline
{}    &  {Age$^{a}$ }  &&&&&{Age$^{b}$}&&&&& {Age$^{c}$}\\
{Galaxy} & {[MgFe]} &  {Fe$_{3}$} &{Mg$_{b}$} &{Ca4227} &{CN2}  &
{[MgFe]} &  {Fe$_{3}$} &{Mg$_{b}$} &{Ca4227} &{CN2}  & \\
  \hline
M32&3.8$^{-0.4}_{+1.2}$&3.4&4.5&3.8&3.6&3.1$^{-0.2}_{+0.3}$&3.0&3.1&3.1&3.0& \\
&&&&&&&&&&&\\
NGC4387&12.9$^{-1.4}_{+1.8}$&13.4&11.1&15.5&13.4&  14.5$^{-1.9}_{+1.9}$&14.5&14.8&14.3&14.5& 12.9$^{-2.2}_{+5.5}$\\
&&&&&&&&&&&\\
NGC4464&15.3$^{-1.8}_{+2.4}$&$>$17.8&10.5&$>$17.8&$>$17.8&$>$17.8&$>$17.8&$>$17.8&14.3&   $>$ 17.8 &18.5$^{-1.6}_{+11.5}$\\
&&&&&&&&&&&\\
NGC4478&5.8$^{-1.0}_{+1.2}$&7.2&3.7&7.8&5.5&   8.2$^{-1.3}_{+1.3}$&8.7&7.4&8.7&8.2&    8.6$^{-1.3}_{+2.4}$\\
&&&&&&&&&&&\\
NGC4473&10.0$^{-0.8}_{+0.4}$&12.3&6.0&17.2&4.6&  11.1$^{-0.5}_{+0.8}$&13.2&10.0&13.2&10.1&   9.9$^{-1.4}_{+1.9}$\\
&&&&&&&&&&&\\
NGC4365&11.1$^{-0.4}_{+1.1}$&17.0&8.3&$>$17.8&10.4& 13.8$^{-0.9}_{+1.2}$&$>$17.8&10.6&$>$17.8&11.8&20.0$^{-7.2}_{+10.0}$\\
&&&&&&&&&&&\\
NGC4621&10.5$^{-0.2}_{+0.3}$&16.2&7.0&$>$17.8&8.4&  10.6$^{-0.2}_{+0.3}$&14.3&9.5&14.1&9.9& 10.6$^{-0.3}_{+0.4}$\\
&&&&&&&&&&&\\
\hline

\end{tabular}
\end{minipage}
\end{table*}


\section*{Acknowledgments} The authors would like to thank P. Coelho and B.
Barbuy for their help for implementing their stellar library, N. Cardiel for his
help for simulating error computations and M. Beasley, J. Cenarro and J.
Falc\'on-Barroso for very  useful suggestions and discussions. The authors thank
the referee for relevant suggestions that improved the original version of the
paper. JLC is a FPU PhD student and  AV is a Ram\'on y Cajal Fellow of the
Spanish Ministry of Education and Science.  This work has been supported  by the
Spanish Ministry of Education and Science grants \emph{AYA2004-03059},
\,\emph{AYA2005-04149} and \emph{AYA2007-67752-C03-01}.


\appendix
\section[]{INCREASING THE AGE SENSITIVITY OF H$_{\beta_{LICK}}$ }

As a consecuence of the characterization of the standard H$_{\beta_{LICK}}$
index definition as a function of wavelength shifts (see section 4.2), 
we have shown how the H$_{\beta_{LICK}}$ index
increases its ability to disentagle ages if all the bandpasses
are shifted 1.7\,{\AA} blueward. In this apendix we provide the main
properties of this alternative index definition and show the most important
differences with respect to the standard, i.e. not shifted, H$_{\beta_{LICK}}$
index definition.

\begin{figure}
\includegraphics[angle=0]{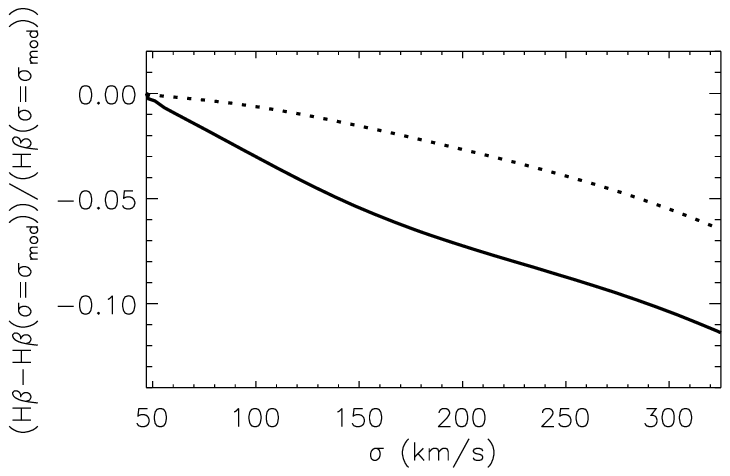}
\caption{Evolution of H$_{\beta_{LICK,o}}$  versus spectral resolution
($\sigma$). The line represents  the mean values for SSPs older than 1\,Gyr.} 
\label{apfigure1}
\end{figure}

\begin{figure}
\includegraphics[angle=0]{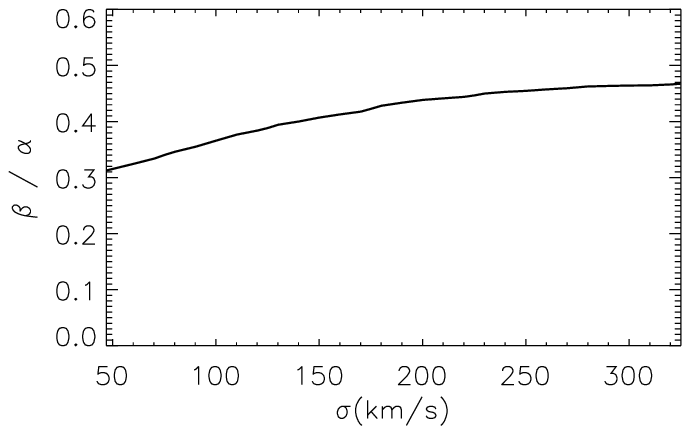}
\caption{Generalized Worthey's parameter, $\beta/\alpha$, versus model spectral
resolution ($\sigma$).  }
\label{apfigure2}
\end{figure}

Table \ref{aptable1} lists the limiting wavelegth of the optimized version of
this index definition (hereafter H$_{\beta_{LICK,o}}$).
The main characteristics are compiled in Table \ref{aptable2}. 
H$_{\beta_{LICK,o}}$ shows a slightly larger age disentangling power than 
H$_{\beta_{LICK}}$ (see Table 2). On the other hand, H$_{\beta_{LICK,o}}$ is
less stable against velocity dispersion than H$_{\beta_{LICK}}$ is. Note,
however, that this dependance, tends to flatten for $\sigma {\rm >
150\,km\,s^{-1}}$ (see Figure
\ref{apfigure1}). Figure \ref{apfigure2} shows that the slightly larger age
resolving power is maintained as function of $\sigma$.

\begin{figure}
\includegraphics[angle=0]{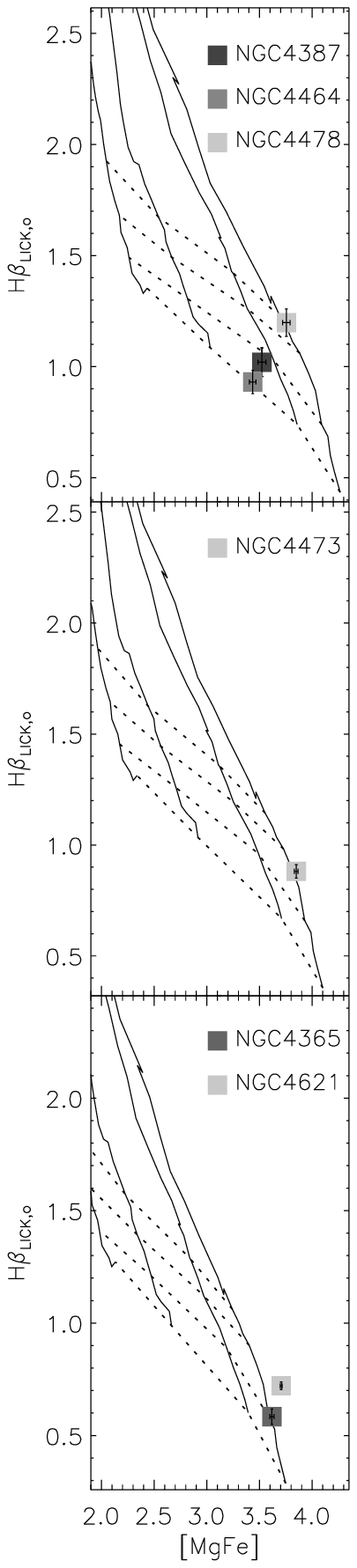}
\caption{[Mg/Fe] vs. H$_{\beta_{LICK,o}}$. Top, middle and bottom panels are for
galaxies with $\sigma_{group}$ = 135, 180, and 300 ${\rm km\,s^{-1}}$,
respectively. Model grids with various ages (dotted lines) and metallicities
(solid lines) are overplotted; the age increases from top to bottom (5.6, 8.0,
11.2 and 17.8 \,Gyr), whereas the metallicity increases from left to right
([Fe/H]=-0.7, -0.4, 0.0 and +0.2).}
\label{apfigure3}
\end{figure}

Figure \ref{apfigure3} shows an age/metallicity diagnostic diagram based on
H$_{\beta_{LICK,o}}$. The plot shows similar age values as those obtained with
which provides similar ages as those obtained with H$_{\beta_{LICK}}$.

\begin{table*}
 \centering
 \begin{minipage}{140mm}
  \caption{H$_{\beta_{LICK}}$ INDEX CONFIGURATION OF BANDPASSES }\label{aptable1}
  \begin{tabular}{@{}lrrrrr@{}}
  \hline
{}    &  {Blue pseudocontinuum}   &
{Feature}  &
{Red pseudocontinuum}&{}&{}  \\
\cline{2-6} \\
{Index} & {\AA} & {\AA} &
 {\AA}&{c$_{1}$}& {c$_{2}$} \\
  \hline
 H$_{\beta_{LICK,o}}$& 4826.175  4846.175  & 4846.175  4874.925&  4874.925 
4889.925&8.590& 0.2439\\

\hline
\end{tabular}
\end{minipage}
\end{table*}

\begin{table*}
\centering
\begin{minipage}{140mm}
 \caption{INDEX CHARACTERISTICS AND UNCERTAINTIES} \label{aptable2}

 \begin{tabular}{@{}lr@{}}
 \hline
{Parameter}      &
{H$_{\beta_{LICK,o}}$}  \\
 \hline
$\frac{\beta}{\alpha}$ &  0.3\\
\\
$\Sigma$ ($10^{-4}$ ${\rm km\,s^{-1}}$)&2.9\\
Index variation by ${\rm \pm 10\,km\,s^{-1}}$ in $\sigma$ (\%)  & 0.29 \\
Maximum stability $\sigma$ range& ${\rm \pm 100\,km\,s^{-1}}$\\
\\
Age uncertainty caused by a $\Delta\lambda$ = 1.5 {\AA} &5.7-18
\,Gyr\\
Maximum $\Delta \lambda$ shifts ($\lambda$, z, rotation curve) compatible with a $+/-$ 2.5 \,Gyr error  ({\AA}) & 0.85\\
\\
(H$_{\beta}$(flux calibrated response curve) - H$_{\beta}$(continuum
removed))/H$_{\beta}$(f.c.r.c.)  & $\lq$ 0.01 \\
\\
S/N (per\,{\AA}) required to distinguish 2.5 \,Gyr at 10 \,Gyr  & $\sim$ 50\\
Minimum age uncertainity for S/N (per\,{\AA}) $>$ 250  & 4.9 \,Gyr \\
\\
Index Sensitivity to [$\alpha$/Fe] variations ([$\Lambda_{CO}$, $\Lambda_{MU}$]) & [0.12,0.23]\\

\hline

\end{tabular}
\end{minipage}
\end{table*}

\bsp

\label{lastpage}

\end{document}